\pdfoutput=1
\documentclass{aa}
\usepackage{graphicx}
\usepackage{amsmath}
\usepackage{txfonts}
\usepackage[colorlinks=true,linkcolor=blue,citecolor=blue,urlcolor=blue]{hyperref}
\begin{document}
\nocite{*} 
    \title{A study of particle acceleration, heating, power deposition, and the damping length of kinetic Alfv\'{e}n waves in non-Maxwellian coronal plasma}

   \author{S. Ayaz\inst{1}\thanks{sa0173@uah.edu/syedayaz263@gmail.com},
          Gary P. Zank\inst{1},
          Imran A. Khan\inst{2,3},
          G. Li\inst{4},
            \and
          Yeimy J. Rivera\inst{5}
          }

   \institute{Department of Space Science and CSPAR, University of Alabama in Huntsville, Huntsville, AL 35899, USA\\
   \and
      Department of Space Sciences, Institute of Space Technology, Islamabad 44000, Pakistan\\
   \and
   Space and Astrophysics Research Lab (SARL), National Center of GIS and Space Applications (NCGSA), Islamabad 44000, Pakistan\\
   \and
   General Linear Space Plasma Lab LLC,4 Foster City, CA 94404, USA\\
           \and
             Center for Astrophysics, Harvard and Smithsonian, Cambridge, MA 02138, USA\\
                          }

   \date{Received: September 26, 2024; Accepted: November 22, 2024}
 
  \abstract
   {The heating of the solar corona and solar wind, particularly through suprathermal particles and kinetic Alfv\'{e}n waves (KAWs) within the 0 – 10 R\textsubscript{Sun} range, has been a subject of great interest for many decades. This study investigates and explores the acceleration and heating of charged particles and the role of KAWs in the solar corona.}
   {We investigate how KAWs transport energy and accelerate and heat the charged particles, focusing on the behavior of perturbed electromagnetic (EM) fields, the Poynting flux vectors, net power transfer through the solar flux loop tubes, resonant particles' speed, group speed, and the damping length of KAWs. The study examines how these elements are influenced by suprathermal particles $(\kappa)$ and the electron-to-ion temperature ratios ($\mathrm{T_e/T_i}$).}
   {We used kinetic plasma theory coupled with the Vlasov-Maxwell model to investigate the dynamics of KAWs and particles. We assumed a collisionless, homogeneous, and low-beta electron-ion plasma in which Alfv\'{e}n waves travel in the kinetic limits; that is, $m_e/m_i \ll \beta \ll 1$. Furthermore, the plasma incorporates suprathermal high-energy particles, necessitating an appropriate distribution function to accurately describe the system. We adopted the Kappa distribution function as the most suitable choice for our analysis.}
   {The results show that the perturbed EM fields are significantly influenced by $\kappa$ and the effect of $\mathrm{T_e/T_i}$. We evaluate both the parallel and perpendicular Poynting fluxes and find that the parallel Poynting flux ($\mathrm{S_z}$) dissipates gradually for lower $\kappa$ values. In contrast, the perpendicular flux ($\mathrm{S_x}$) dissipates quickly over shorter distances. Power deposition in solar flux tubes is significantly influenced by $\kappa$ and $\mathrm{T_e/T_i}$. We find that particles can heat the solar corona over long distances (R\textsubscript{Sun}) in the parallel direction and short distances in the perpendicular direction. The group velocity of KAWs increases for lower $\kappa$ values, and the damping length, $\mathrm{L_G}$, is enhanced under lower $\kappa$, suggesting longer energy transport distances (R\textsubscript{Sun}). These findings offer a comprehensive understanding of particle-wave interactions in the solar corona and wind, with potential applications for missions such as the Parker Solar Probe, (PSP), and can also apply to other environments where non-Maxwellian particle distributions are frequently observed.}
   {}

               \maketitle
               %
\section{Introduction}

In space plasma environments such as the solar wind, solar corona, planetary magnetospheres, and the interstellar medium, turbulence plays a critical role in the dissipation of energy and charged particle acceleration \citep{jokipii1966cosmic, bieber1996dominant, farmer2004wave, engelbrecht2013ab}. The dissipation and particle acceleration processes are central to understanding how energy is transferred across different scales, from large fluid-like motions down to smaller kinetic scales. The dissipation of turbulent energy helps to explain key phenomena, including the nonadiabatic temperature profiles observed in the solar wind and planetary magnetospheres \citep[]{williams1995dissipation,matthaeus1999coronal, adhikari2015transport, adhikari2015turbulence}, which increase with distance from the Sun and the central planet, respectively \citep[]{richardson1995radial, saur2004turbulent, bagenal2011flow,von2014turbulent}. Among the various scaling mechanisms observed in different space plasma environments, the Kolmogorov scaling ($k^{-5/3}$) \citep[]{matthaeus1982evaluation, bale2005measurement}, ion scales (such as the ion Larmor radius and ion inertial length) \citep[]{leamon1999dissipation, chen2014ion}, sub-ion range \citep{alexandrova2009universality, chen2010anisotropy}, and electron range \citep[]{denskat1984interplanetary}, to name a few, are particularly important. Although observations at these small scales are relatively scarce and often subject to varying interpretations, they provide crucial insights into the turbulent fluctuations within the dissipation range \citep[]{denskat1984interplanetary}. At scales comparable to the ion Larmor radius, Alfv\'{e}n waves are thought to be slightly damped, indicating a transition to the dissipation range. \cite{schreiner2017model} presented an analytical dissipation model for solar wind turbulence at electron scales, combining energy transport from large to small scales with collisionless damping. The model suggests that wave-particle interactions involving kinetic Alfv\'{e}n waves (KAWs) are the primary damping mechanism. Our research further supports this by analytically demonstrating that Alfv\'{e}n waves dissipate more rapidly at very small, kinetic scales. Over time, interest in the heating effects of Alfv\'{e}n waves in the solar corona has grown, with numerous studies underscoring their significance (for details, see \cite[]{matthaeus1999coronal, zank2018theory,zank2021turbulence, yalim2024mixing} and references therein).

Despite advances, our understanding of the kinetic processes responsible for turbulent dissipation, wave heating, and energy transfer remains incomplete. This knowledge is essential for explaining phenomena such as solar wind acceleration and coronal heating \citep{parker1958dynamics, richardson1995radial, fox2016solar}, as well as analogous astrophysical processes. More recently, \cite{bowen2024extended} investigated cyclotron resonant heating of the turbulent solar wind at ion-kinetic scales, finding that ion cyclotron heating rates correlate strongly with the turbulent energy cascade rate. This correlation implies that cyclotron heating is an important dissipation mechanism in the solar wind.

The several promising frequently used dissipation mechanisms of Alfv\'{e}n waves are: the resonant heating in coronal loops \citep{hollweg1978alfven, hollweg1981alfven, schwartz1984chromospheric}, turbulent heating due to Kelvin-Helmholtz instabilities \citep{heyvaerts1983coronal}, turbulence cascade \citep{williams1995dissipation, matthaeus1999coronal}, phase mixing caused by Alfv\'{e}n velocity gradients \citep{heyvaerts1983coronal, abdelatif1987heating}, spatial resonance -- that is, resonant mode conversion \citep{ionson1978resonant, lee1986behavior}, nonlinear mode coupling \citep{chin1972nonlinear, uchida1974excess, wentzel1974coronal} -- and kinetic Landau damping \citep{stefant1970alfven, hollweg1971density}, to name a few. In a microphysical context, the competition among these mechanisms is largely influenced by the presence of smaller-scale Alfv\'{e}n waves. Generally, the dissipation rate of Alfv\'{e}n waves is proportional to the square of the wave number ($k^2$). For the heating rate resulting from Alfv\'{e}n wave dissipation to balance the energy loss rate in coronal plasmas, these
small-scale Alfv\'{e}n waves (i.e., KAWs) must have scales close to the particle kinetic scales typical of coronal plasmas, such as the ion gyroradius ($\rho_i$) or the electron inertial length ($\lambda_e$), both of which are significantly smaller than the mean free path of coronal plasmas. At these particle kinetic scales, Alfv\'{e}n waves transition into dispersive KAWs. The wave-particle interactions, particularly through Landau damping, become critically significant in the context of kinetic dissipation of wave energy and plasma particle heating. In this regard, one of the most promising candidates for a coronal heating mechanism is the kinetic dissipation of KAWs via wave-particle interactions through the Landau damping mechanism. This approach offers a comprehensive framework for applying KAWs to the coronal heating problem, including chromospheric heating and solar wind heating. Consequently, we utilize this mechanism within the framework of the most generalized Vlasov-Maxwellian model, situated in the domain of kinetic theory in plasma physics.

The first detection of Alfv\'{e}n waves \citep{alfven1942existence} in the solar corona, as was reported by \cite{tomczyk2007alfven}, opened up a new avenue of research in space physics. However, their observational evidence suggested that these waves are weak and might not significantly energize the solar wind and corona. \cite{de2007chromospheric}, followed  \cite{tomczyk2007alfven}, used 3D radiative MHD simulations and found that Alfv\'{e}n waves are strong enough to accelerate and heat the solar wind and the solar corona. These waves are thought to play a crucial role in the heating mechanisms of the solar corona \citep{del2002coronal,khan2020solar}. \cite{davila1987heating} explored solar corona heating through the resonant absorption of Alfv\'{e}n waves. He suggested that this method is a viable mechanism for heating the corona of the Sun and other late-type stars. Alfv\'{e}n waves can energize the solar wind through the work done by the Alfv\'{e}n waves pressure force, which directly accelerates plasma away from the Sun \citep{hollweg1973alfven,wang1993flux}. The significant interest in the heating phenomena of Alfv\'{e}n waves in the solar corona has evolved over time, highlighted by various studies (e.g., \citep{matthaeus1999coronal, zank2018theory, zank2021turbulence}). Several spacecraft observations, including in situ measurements, have demonstrated the presence of large-amplitude outward-propagating Alfv\'{e}n waves in the interplanetary medium \citep{belcher1971large,tu1995mhd,bruno2013solar}. Moreover, remote observations have identified Alfv\'{e}n wave-like motions in the low corona that carry an energy flux sufficient to power the solar wind \citep{de2007chromospheric}. Several considerations suggest that much of this heating results from the dissipation of MHD "wave/turbulence-driven” (W/T) models or the “reconnection/loop-opening” (RLO) models \citep{cranmer2010can}, further explored by \cite{zank2018theory,zank2021turbulence}. Alfv\'{e}n waves have interested scientists for many decades due to their remarkable role in transporting energy across space and astrophysical environments (see \cite{matthaeus1999coronal,cramer2011physics,zank2018theory, wu2020kinetic, zank2023linear, zank2024characterization}, and references therein). These waves have practical applications in laboratory plasma, particularly in transferring energy within fusion devices \citep{hasegawa1982alfven}.

Observations have confirmed the presence of short-wavelength Alfv\'{e}n waves (also known as KAWs) and electron beams, as has been demonstrated by \cite{kerr2016simulations} using the RAdiative hydro DYNamics code (RADYN). Their work synthesized various chromospheric spectral lines, revealing an intriguing distinction: Mg II profiles in wave-heated simulations differed from those in electron beam simulations, with the wave-heated model more closely matching \textit{IRIS} observations. This suggests that detailed studies of spectral lines formed in the mid to upper chromosphere may help distinguish between heating mechanisms. The high-resolution in situ measurements of electromagnetic (EM) fluctuations and plasma distribution functions from satellites, including the Magnetospheric Multiscale (MMS) mission, Cluster, and the PSP, have created unprecedented opportunities to investigate the complex plasma dynamics within the sub-ion scale of turbulence (for details, see \cite{sahraoui2009evidence}, \cite{alexandrova2009universality, alexandrova2012solar, alexandrova2013solar}, \cite{chen2019evidence}, and references therein). The primary aim of these studies is to unravel the mechanisms of energy dissipation in weakly collisional plasmas and to understand how electrons and ions gain energy in these environments \citep{parashar2015transition}. \cite{hui1992electron} conducted hybrid simulations showing that large-amplitude, short-wavelength Alfvén waves can accelerate electrons to velocities exceeding the Alfvén speed. Observational support for small perpendicular-wavelength Alfv\'{e}n waves has been obtained through sounding rocket missions \citep{boehm1990high} and the Freja satellite \citep{Wahlund1994, Louarn1994, boehm1995freja}.

Very recently, in situ measurements by \cite{rivera2024situ} using the PSP and Solar Orbiter spacecraft have provided some of the most compelling evidence yet on Alfv\'{e}n wave heating in the solar corona and solar wind. These observations reveal that after leaving the Sun’s corona, the solar wind continues to accelerate and cool at a slower rate than expected for a freely expanding adiabatic gas. Alfv\'{e}n waves, which are perturbations in the interplanetary magnetic field, transport energy and are shown to induce both heating and acceleration in the plasma between the corona’s outer edge and near Venus’s orbit. The study indicates that the damping and mechanical work performed by these large-amplitude Alfv\'{e}n waves are sufficient to power the heating and acceleration of the fast solar wind in the inner heliosphere. Early analyses by \cite{chaston2014heating} estimated the heating rate of KAWs and found that the heating occurs primarily in the perpendicular direction. Observations from various spacecraft also confirm that KAWs dissipate and contribute to plasma heating during propagation \citep{Wygant2002, Lysak2003, gershman2017}, highlighting their essential role in transport, heating, and acceleration processes in space and astrophysical plasmas. The high-resolution imaging and spectroscopy of flare ribbons provided by the Daniel K. Inouye Solar Telescope (DKIST) are crucial for understanding energy transport in solar flares. According to 1D radiative hydrodynamics simulations, the lower solar atmosphere responds distinctly to heating by Alfv\'{e}nic waves compared to heating by electron beams, though there are notable similarities as well \citep{kerr2016simulations, reep2016alfvenic, reep2018hydrodynamic}. Localized electric fields associated with KAWs are often observed to accelerate electrons, forming electron beams (e.g., \cite{artemyev2015acceleration}; \cite{damiano2015ion, damiano2016ion}). Electron beams are ubiquitous in space and astrophysical plasmas, detected remotely through radio observations and in situ by numerous satellites. Examples include beams associated with solar bursts, Earth's foreshocks, and Earth's magnetosphere. In the context of our model, which explores KAWs in a kappa-distributed plasma via kinetic plasma theory, we focus on the ion gyro-radius and the electron beam aligned with the electric field. This configuration can indeed excite and heat the coronal region, highlighting the strong connection between KAW-driven electron beams and particle acceleration activity in such plasma environments.

\cite{hasegawa1975kinetic} and \cite{hasegawa1977kinetic} were the first to study Alfv\'{e}n waves by incorporating the effect of a finite perpendicular wavelength -- also known as gyroradius corrections -- leading to the identification of kinetic Alfv\'{e}n waves (KAWs). \cite{hasegawa1975kinetic} added gyroradius correction terms and derived the dispersion relation for obliquely propagating Alfv\'{e}n waves, now recognized as the dispersion relation for KAWs. They utilized the Vlasov model and assumed a Maxwellian distribution function to derive the following dispersion relation for KAWs:
\begin{equation*}
    \omega^2=k_{\parallel}^2 \mathrm{v_{A}}^2\Bigr(1+\frac{3}{4}k_{\perp}^2 \rho_{i}^2+\mathrm{\frac{T_e}{T_i}} k_{\perp}^2 \rho_{i}^2\Bigr),
\end{equation*}
where $k_{\perp}$ and $k_{\parallel}$ are the perpendicular and parallel wavevectors, the symbols, $\perp$ and $\parallel$, denote the directions perpendicular and parallel to the mean/ambient magnetic field $\mathrm{B_0}$, $\rho_i$ is the ion gyroradius, and $\mathrm{v_A}$ is the Alfv\'{e}n speed, respectively. In the expression, the first term $\frac{3}{4} k_{\perp}^2 \rho_{i}^2$ is the gyro-radius correction, and the second term $\mathrm{\frac{T_e}{T_i}} k_{\perp}^2 \rho_{i}^2$ arises due to the wave's obliqueness. Following Hasegawa's pioneering work, KAWs garnered significant attention. \cite{lysak1996kinetic} further explored KAWs by incorporating the full kinetic effects of electrons and ions, rather than just the gyroradius correction terms. They demonstrated that the parallel phase velocity ($\omega/k_{\parallel}$) of these waves exceeds the thermal speed of electrons in the presence of hot ions \citep{vstverak2009radial}.

\cite{naim2014drift} extended \cite{lysak1996kinetic}'s work by considering density inhomogeneity in the bi-Maxwellian distribution function to study KAWs instability. They found that density inhomogeneity destabilizes KAWs, but temperature anisotropy can stabilize them. KAWs have been extensively studied using various velocity distributions within kinetic plasma theory \citep{shukla2009study,vranjes2010kinetic,liu2014dispersion,khan2019distinct,ayaz2020alfven,wu2020kinetic,barik2020nonresonant,barik2021kinetic}. In our previous work \citep{ayaz2019dispersion}, we examined the dispersion and damping relations of KAWs in the Cairns-distributed plasmas, finding that nonthermal particles significantly influence the real phase velocity and damping rate of KAWs. \cite{khan2019distinct,khan2020solar} studied KAWs in kappa and non-extensive velocity distribution functions, finding that suprathermal particles and non-extensive parameters, characterized by the index $\kappa$ and $q$, significantly affect the waves' characteristics, enhancing their heating and energy transport over extended spatial distances in space.

KAWs have been extensively investigated by \cite{barik2020nonresonant,barik2023kinetic} in space plasmas with $\kappa-$electrons, finding that the ion beam velocity can potentially excite the growth rate of KAWs. More recently, they examined the excitation of KAWs by multiple free energy sources, concluding that a combination of positive velocity shear with counter-streaming beam ions or parallel streaming beam ions with negative velocity shear favors KAW excitation \citep{barik2021kinetic}. KAWs are likely to play a significant role in particle acceleration and energy transporting to the coronal plasma in the form of heat \citep{Khan2019a}. These waves are considered one of the primary agents responsible for transporting energy from the distant magnetosphere to the auroral ionosphere 
\citep{louarn1994observation,seyler1995theory,chaston2005drift,chaston2006ionospheric,ergun2005localized,artemyev2015electron,gershman2017}, in particle energization at the formation of depolarization fronts \citep{ganguli2018kinetic,ukhorskiy2022cross}, and in electron trapping and acceleration in solar flares \citep{artemyev2016electron}, and the counterpart (see \cite{che2023particle}).

KAWs have garnered significant attention over the past few decades due to their potential role in various space environments. They are proposed as an energy source for planetary auroras \citep{chaston2003properties, Keiling2003}  and for driving atmospheric escape induced by space weather \citep{chaston2006ionospheric}. KAWs are key to understanding space and astrophysical turbulence \citep{Boldyrev2012}, potentially solving the plasma heating mystery in the solar corona, solar wind, magnetospheres, and the interstellar medium \citep{Liu2023}. KAWs are the most rapidly damped waves \citep{shukla2009study} that efficiently transfer energy to plasma particles through Landau damping \citep{khan2020solar,ayaz2020alfven}. 

\cite{Temerin2001} provided the first evidence of small-scale KAWs within the plasma sheet and plasma sheet boundary layer (PSBL). These waves likely play a crucial role in the local heating of the plasma sheet, potentially generating earthward-moving, magnetic field-aligned electron beams, and transversely heated ions. Extensive observational studies support the existence of KAWs and their significant role in energy transport across various regions of Earth's magnetosphere, including the magnetopause \citep{Johnson2001, chaston2005drift}, magnetosheath and auroral regions \citep{Boehm1990, Louarn1994, Wahlund1994}, magnetotail and PSBL \citep{Wygant2002, Keiling2003, Duan2012}, and central plasma sheet \citep{Keiling2001}. \cite{Wygant2002}, presented observational evidence showing that the increase in Poynting flux in the PSBL during the substorm expansion phase is related to KAWs. Several researchers have extensively studied these small fluctuating waves 
\citep{Lee1994, shukla2009study, wu2020kinetic, khan2020solar, Liu2023, barik2023kinetic, Ayaz2024}, highlighted their significant role in energy transport and particle acceleration in space plasmas.

In this proposed research study, we aim to address the role of perturbed EM fields in identifying KAWs \citep{Lysak2003, Khan2019a}. In space plasmas, if the ratio of the perturbed EM fields equals Alfv\'{e}n speed -- the KAWs are most likely to be present in those regions. These fluctuating EM fields are crucial for energy transport in solar wind turbulence \citep{Narita2020}. \cite{Narita2020} revisited KAWs analytically, deriving transport ratios and scaling laws for the electric and magnetic fields to aid in wave mode identification in spacecraft observations. This is particularly important for missions like the PSP, Solar Orbiter, and BepiColombo’s cruise to Mercury.

KAWs -- defined by the condition where the perpendicular spatial scale of Alfv\'{e}n wave approaches the ion gyroradius ($\rho_i$), can support significant electric and magnetic field fluctuations in the parallel and perpendicular directions \citep{xiang2022mode}. These fluctuations facilitate energy transfer between the wave field and plasma particles via Landau or transit-time interactions \citep{Barnes1966, Hasegawa1976,gershman2017}. Recently, \cite{Xunaira2023} studied the instability and energy transport of KAWs in the solar corona using the field-aligned drift velocity distribution function, finding that the field-aligned drift velocity significantly influenced the EM fields of the KAWs. \cite{khan2019distinct} investigated the perturbations of EM field ratios for different values of the non-extensive parameter $q$. They found nearly similar results observed by polar spacecraft \citep{Wygant2002}. The knowledge of these perturbed EM fields is pivotal in wave-particle interactions through Landau resonance and can be used to understand how KAWs transport energy in the solar corona \citep{Ayaz2024}. 

\cite{Hollweg1999} provides a comprehensive analysis of KAW physics. Many observational studies of the solar wind \citep{Bale2005, Sahraoui2012, Podesta2012, Salem2012, Chen2012, Kiyani2012}, theories
\citep{Howes2008a, Howes2008b, Schekochihin2009}, and numerical simulations \citep{Gary2004, TenBarge2012} have suggested that KAWs play a significant role in the dissipation of turbulent energy. Specifically, understanding the perturbations in EM fields is important for comprehending how KAWs transport energy in space. The EM fields can be further used in the expressions of the Poynting flux vector, which quantifies the energy per unit time per unit area and evolves according to the Poynting theorem \citep{Khan2019a, Ayaz2024}. Essentially, the Poynting flux vector reveals whether a wave can effectively carry or deliver its energy over long distances from its source. \cite{Lysak2003} investigated the Poynting flux of KAWs in a Maxwellian-distributed plasma. Extending this research, \cite{khan2020solar} analyzed KAWs in a more generalized $\kappa$ distribution and discovered that the Poynting flux vector of KAWs is rapidly damped in the presence of fewer suprathermal particles. \cite{Shay2011} examined the Poynting flux of KAWs associated with quadrupolar $\mathrm{B_y}$ using kinetic particle-in-cell (PIC) simulations and Cluster satellite observations, finding that KAWs can carry significant Poynting flux in the magnetosphere.

KAWs can be excited by various mechanisms such as phase mixing \citep{Heyvaerts1983}, nonlinear decay from magnetohydrodynamic Alfv\'{e}n waves \citep{Zhao2011, Zhao2013, Zhao2014}, turbulence cascade \citep{Bian2010, Zhao2013}, electron beams \citep{Chen2012}, magnetic reconnection \citep{Cranmer2018}, and photospheric fluctuations \citep{Ulrich1996}. Near the Sun, these waves start their journey carrying a significant amount of Poynting flux, approximately $10^{5}$ W $m^{-2}$ \citep{Srivastava2017}. Although some energy is lost after partial reflection from the transition region \citep{zank2021turbulence}, the waves can still carry a substantial amount of energy, around $\sim 10^{3}$ W $m^{-2}$ \citep{khan2020solar}, as they travel through the corona. Observations indicate that a Poynting flux of $10^{2}$ - $10^{4}$ W $m^{-2}$ is sufficient to heat the corona to a million degrees Kelvin \citep{Srivastava2017}. These insights into the Poynting flux and energy transport mechanisms of KAWs showcase their significant role in space and astrophysical environments.

KAWs propagate obliquely to the ambient magnetic field $\mathrm{B_0}$ with wavevectors such that $k_{\perp}\gg k_{\parallel}$ \citep{Narita2020,Ayaz2024}. This characteristic means KAWs primarily carry energy along the magnetic field lines \citep{Lysak2003}, as is indicated by the Poynting flux vector ($\mathrm{S_z}$) in the z-axis (i.e., in the parallel direction). Although KAWs also transport some energy in the perpendicular direction ($\mathrm{S_x}$) \citep{khan2020solar}, this is often less significant ($\mathrm{S_x\ll S_z}$) in regions like the aurora and plasma sheet boundary layers \citep{Lysak2003, khan2019distinct, lysak2023kinetic}. In this study, we focus on KAWs in the solar coronal regions, particularly within solar flux tube loops, where both perpendicular and parallel energy transformations and heating are critical. More recently, we investigated the parallel and perpendicular Poynting fluxes of KAWs in the solar corona \citep{Ayaz2024}. We found that the perpendicular Poynting vector ($\mathrm{S_x}$) for KAWs dissipates rapidly, leading to plasma heating over short distances. The importance of perpendicular heating is also highlighted by \cite{zank2018theory}, who investigated the heating of core solar coronal protons using a turbulence transport model and found significant perpendicular heating within 0.3 R\textsubscript{Sun}. In this research, we utilize both the perpendicular and parallel Poynting vectors ($\mathrm{S_x}$ and $\mathrm{S_z}$) to examine the total power transfer rate of KAWs in solar flux tube loops. This comprehensive approach helps to understand how KAWs contribute to energy transportation and heating in the solar corona.

In general, KAWs transfer energy to the plasma, causing variations in energy distribution as the wave propagates through space \citep{khan2020solar}. This energy transfer significantly affects the behavior of charged particles, including their acceleration and speed, which remains a dynamic area of research. However, by employing a kinetic Vlasov-Maxwell model, we derived the net velocity expressions of the particles, providing a comprehensive understanding of their behavior and interactions within the plasma. Moreover, \cite{tiwari2008shear} studied the growth and damping length of KAWs in the PSBL using the loss-cone distribution function. Their findings indicate that the growth length of KAWs decreases with varying values of the loss cone parameter $\mathrm{J\geq 0}$. This suggests that more energetic particles are available to transfer energy to the waves through resonance wave-particle interactions. We extended this approach to a $\kappa-$distributed plasma to investigate how suprathermal particles influence the damping length of KAWs. These insights are pivotal for understanding the dynamics of KAWs and their role in energy transport and particle acceleration in various space plasma regions.

In this manuscript, we aim to address several key aspects of induced heating and acceleration of charged particles within the solar corona. These include:

(a). The perturbed EM field ratios (imaginary parts).

(b). The Poynting flux vectors of KAWs.

(c). The total power rate of KAWs in the solar flux tube loop.

(d). The net resonance velocity of the particles.

(e). The group velocity of KAWs.

(f). The characteristic damping length of KAWs.

We investigate these issues (a) - (f) using kinetic plasma theory in Kappa-distributed plasmas. Our analysis considers the influence of suprathermal particles characterized by the parameter $\kappa$, along with variations in the electron-to-ion temperature ratio ($\mathrm{T_e/T_i}$) and the height (h) relative to the Sun's radius (R\textsubscript{Sun}). This comprehensive approach helps us understand how these elements affect the behavior and dynamics of KAWs and particles in the solar corona.

The manuscript is organized as follows. Section 2 provides the KAWs model that supports our analytical framework. Section 3 provides an in-depth numerical analysis to validate our theoretical results. Finally, Section 4 offers a comprehensive discussion and conclusion, summarizing our key findings and their broad implications.

\section{Kinetic Alfv\'{e}n wave model}

We start with the dispersion relation described by \cite{lysak1996kinetic, lysak1998relationship} for a collisionless, homogeneous, and low-$\beta$ plasma (i.e., the ratio of thermal to magnetic pressure $\beta=\mathrm{P/(B^2}/2\mu_0)\ll1$ \citep{zank2022turbulence}):
 \begin{equation}
\left( 
\begin{array}{ccc}
\epsilon _{xx}-\frac{k_{\Vert }^{2}c^{2}}{\omega ^{2}} &  & \frac{k_{\bot
}k_{\Vert }c^{2}}{\omega ^{2}} \\ 
&  &  \\ 
\frac{k_{\bot }k_{\Vert }c^{2}}{\omega ^{2}} &  & \epsilon _{zz}-\frac{%
k_{\bot }^{2}c^{2}}{\omega ^{2}}%
\end{array}%
\right) \left( 
\begin{array}{c}
\mathrm{E_{x}} \\ 
\\ 
\mathrm{E_{z}}%
\end{array}%
\right)=0 
\label{eq:1}.
\end{equation}
In Eq. (\ref{eq:1}), $\omega$ is assumed to be a complex quantity -- that is, $\omega=\omega_r+i\omega_i$ with $\omega_i\ll\omega_r$ -- and the subscripts $r$ and $i$ denote the real and imaginary, respectively. The term $ck_{\perp,\parallel}/\omega$ is the perpendicular/parallel refractive indices, $\epsilon_{xx,zz}$ is the x,z-component of the permittivity tensors, and $c$ is the speed of light.

In developing our model, specifically in deriving the expressions for the two permittivity tensors ($\epsilon_{xx}$ and $\epsilon_{zz}$) given in \hyperref[Appendix: A]{\textcolor{blue}{Appendix A}}, we considered a constant mean/ambient magnetic field oriented along the z axis. This assumption ensures that the system remains homogeneous, so we have no inhomogeneity within our model. In the model, we assumed the magnetic field values, \textbf{B}, could range between 10 and 100 G, reflecting different regions of the solar corona at various heights or distances from the Sun (R\textsubscript{Sun}). The specific value chosen for our model is simply a numerical selection from this range; however, it remains constant throughout our analysis. On the other hand, if we were to account for inhomogeneity in the system, the whole model, permittivity tensors ($\epsilon_{xx}$ and $\epsilon_{zz}$) would require substantial modifications that are beyond the scope of the present study. Currently, we are uncertain whether introducing inhomogeneity would lead to significant findings.

\begin{figure}[htp]
        \centering
       {
            \includegraphics[width=.47\linewidth]{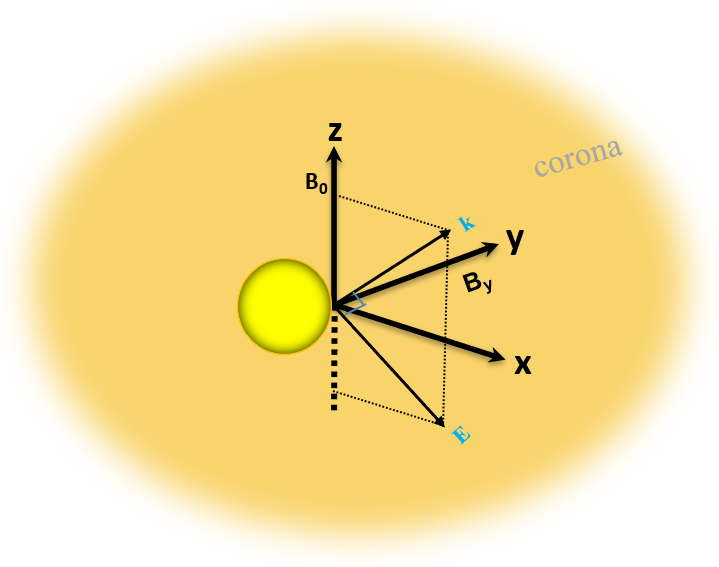}
                    }\quad
        {
            \includegraphics[width=.47\linewidth]{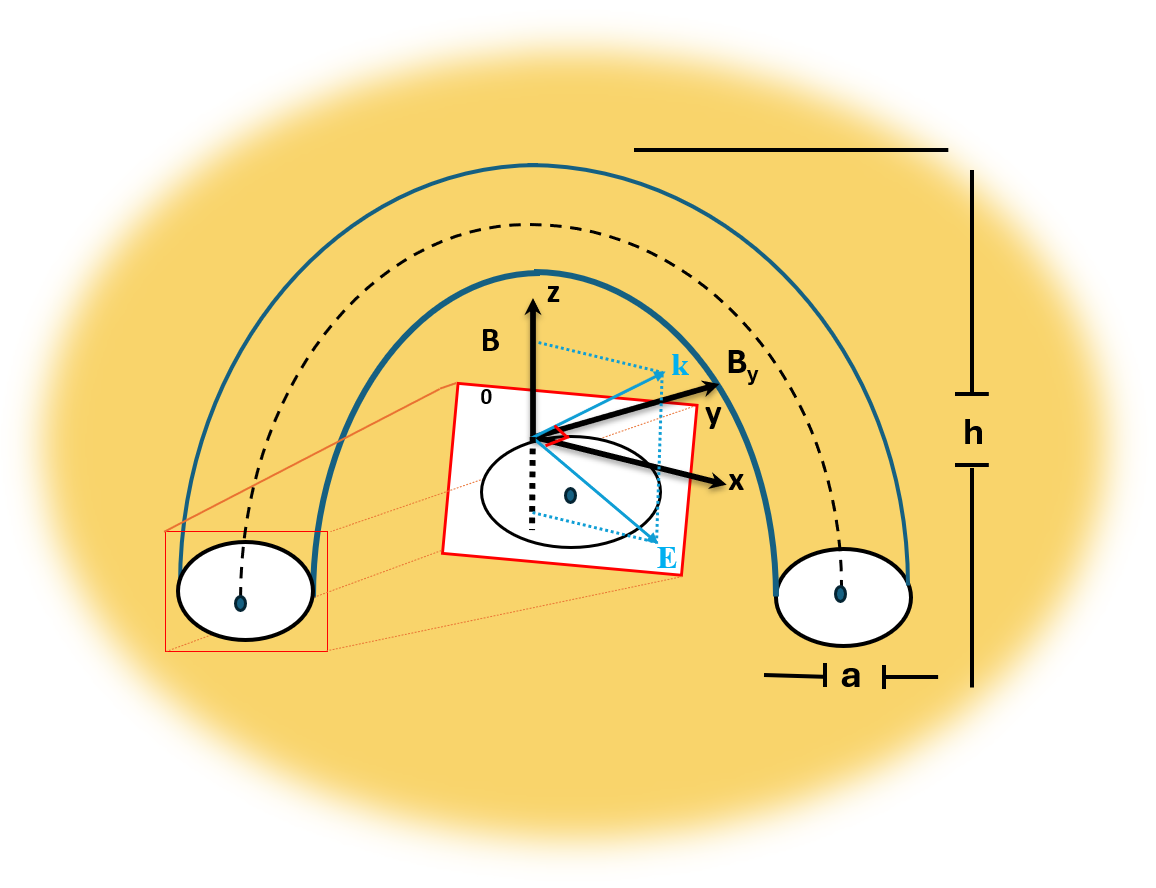}
                    }
        \caption{Geometry of obliquely propagating KAWs. The left panel depicts the generation of the waves (KAWs) somewhere near the Sun's surface and propagates into the solar corona. The right panel shows a detailed schematic of KAWs within a solar flux tube loop. The flux tube, with a height, $\mathrm{h}$, and a circular crosssection of radius $a$, provides a structured pathway for wave propagation. This fitted geometry highlights how KAWs navigate through the solar atmosphere, emphasizing the importance of their spatial characteristics in understanding energy distribution and particle acceleration in the solar corona.}
    \label{A}
\end{figure}

Moreover, we assumed that the plasma is supporting low-frequency Alfv\'{e}n waves, which means that the wave frequency is less than the ion gyrofrequency ($\omega\gg\Omega_i$). The wave propagates in a cylindrical geometry where the ambient/mean magnetic field $\mathrm{B_0}$ points in the z direction. The wave magnetic field, \textbf{B}, is in the y direction and the perturbed electric field, \textbf{E}, and the wave vector, \textbf{k}, are in the x-z plane (Fig. \ref{A}). Specifically, we are interested in the solar coronal region, however, we fit this obliquely propagating wave's geometry in the solar flux tube loop, as is shown in Fig. \ref{A} (right panel).

In addition to the above description, we assumed that the plasma possesses suprathermal high-energy particles. For the sake of simplicity, in this research work, we are considering the isotropic nature of the particles. To model such plasma configuration, we need a suitable velocity distribution function for which we employ the isotropic kappa distribution function of the form \citep{summers1991modified}:
\begin{equation}
    f_{0s}(v)=\frac{1}{\pi^{3/2}\mathrm{v_{Ts}^3}}\frac{\Gamma(\kappa+1)}{\Gamma(\kappa-1/2)}
    \Bigr(1+\frac{1}{\kappa}\frac{v^2}{ \mathrm{v_{Ts}^2}}\Bigr)^{-\kappa-1} 
    \label{eq:2}.
\end{equation}
In the distribution function, the symbol $\Gamma$ denotes the gamma function, $\mathrm{v_{Ts}}=\sqrt{\frac{2\mathrm{T}_s}{m_s}}$ is the thermal speed of $s$-species ($s$ is ions or electrons) of mass $m_s$ and temperature $\mathrm{T}_s$. The spectral index $\kappa$ should be greater than $3/2$ and can take any values other than $3/2$. Interestingly, the distribution function is reduced to the Maxwellian distribution when $\kappa \rightarrow \infty$.

Using Eq. (\ref{eq:2}) in the two permittivity tensors ($\epsilon_{xx}$ and $\epsilon_{zz}$) expressions and on substitute $\epsilon_{xx}$ and $\epsilon_{zz}$ in 
\begin{equation*}
    \mathrm{D}=(\epsilon_{xx}-n_{\parallel}^2)\epsilon_{zz}-\epsilon_{xx}n_{\perp}^2,
\end{equation*}
the real and imaginary frequencies of KAWs can be obtained (for instance, see the details in \hyperref[Appendix: A]{\textcolor{blue}{Appendix A}}). These dispersion relations are further used in Faraday's law and employing Eq. (\ref{eq:1}), we get the real and imaginary parts of the perturbed EM field ratios, respectively. The mathematical steps for the derivation of these expressions are provided in the \hyperref[Appendix: A]{\textcolor{blue}{Appendix A}}.

Furthermore, we utilize these EM field expressions and get the parallel and perpendicular Poynting flux vectors ($\mathrm{S_z}$ and $\mathrm{S_x}$) of KAWs and the net power transfer by KAWs through the solar coronal flux loop. \cite{huang2018kinetic} studied KAWs and demonstrated that these waves are capable of efficiently transporting energy and play an important role in facilitating magnetic reconnection. Using particle-in-cell simulations, they examined Hall fields in the magnetic reconnection region and found that (1) the Hall electric field is balanced by the ion pressure gradient and (2) the ratio of Hall electric field to Hall magnetic field is on the order of Alfvén speed. These results are consistent with our current analysis of the perturbed EM field ratios. The Simulation results of \cite{huang2018kinetic} also indicate that KAWs are excited in the reconnection site and then transmitted along the separatrices. The wave Poynting flux propagates parallel to the magnetic field lines, carrying substantial energy. In our work, we also find that KAWs Poynting flux dissipates at a slower rate in the Kappa distribution demonstrating that KAWs carry significant energy along the magnetic field lines (i.e., in the parallel direction).

In general, when KAWs interact with ions and electrons in a resonant region, Landau resonance primarily generates field-aligned electron beams, while cyclotron resonance with ions results in cross-field ion heating. In the present study, we have electron-ion plasma, where KAWs interact with both species via the Landau damping mechanism. In our model, particularly during the derivation of the expressions for the permittivity tensors ($\epsilon_{xx}$ and $\epsilon_{zz}$), we considered zero modes; in other words, $n$ = 0. In this scenario, the species will experience Landau resonance when the condition $\omega= k_{\parallel}\mathrm{v_A}$ is met. It is important to note that pure or Magnetohydrodynamic (MHD) Alfv\'{e}n waves do not experience Landau resonance because they propagate strictly parallel to the ambient magnetic field (as is illustrated in Fig. \ref{A}, the left panel). In the parallel direction, there is no electric field component present, which is necessary for wave-particle interaction to occur. This distinction is crucial for understanding the dynamics of KAWs in electron-ion plasma, where the interaction mechanisms, driven by Landau damping, significantly influence energy transfer and heating processes within the plasma.

For the wave-particle interaction to occur, the wave should be obliquely propagating so that it has a non-vanishing electric field component in the direction of the ambient magnetic field. However, based on the $n$ = 0 assumption in our model, we do not have the cyclotron resonance. On the other hand, in addition to Landau resonance, if other $n \neq 0$ terms are considered; that is,  $n$ = 1  or other greater values of $n$, one can include cyclotron resonance and can get other modes (i.e., R and L waves).

As was stated in the introduction, Alfvén waves transition into dispersive KAWs in the particles' kinetic scale. The wave-particle interactions via Landau damping become important in the context of kinetic dissipation of wave energy and plasma particle heating. The most promising candidate for the coronal heating mechanism is the kinetic dissipation of KAWs via wave-particle interactions through the Landau mechanism. We utilize this Landau mechanism and find the net resonance speed the particle gains from KAWs during the Landau mechanism.

Ultimately, dispersion relations are employed to determine both the group velocity and the characteristic damping length of KAWs. The group velocity quantifies the rate of energy propagation by KAWs, while the damping length reveals the extent to which KAWs transport energy within the solar corona before energy dissipation occurs. Detailed, step-by-step derivations for the Poynting flux vectors, net power deposition, resonance velocity of particles, group velocity, and characteristic damping length of KAWs are provided in \hyperref[Appendix: B]{\textcolor{blue}{Appendix B}}.

\section{Numerical analysis}
\subsection{The perturbed EM-field ratios of KAWs}
We begin by analyzing the normalized imaginary perturbed EM field $\text{Im} \mathrm{(E_x/v_{A}B_{y}})$ 
of KAWs in the solar coronal region. Our interest in this expression (e.g., Eq. (\ref{eq:15})) stems from our focus on a region where even small contributions play a crucial role. These minor contributions are often overlooked in the literature, but our work evaluates this field ratio for various values of the index parameter $\kappa$ and the ratio of electron-to-ion temperature $\mathrm{T_e/T_i}$.

\begin{figure}[h!]
        \centering
       {
            \includegraphics[width=.47\linewidth]{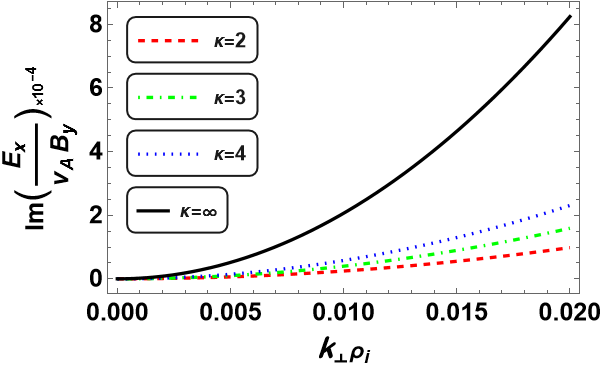}
                    }\quad
        {
            \includegraphics[width=.47\linewidth]{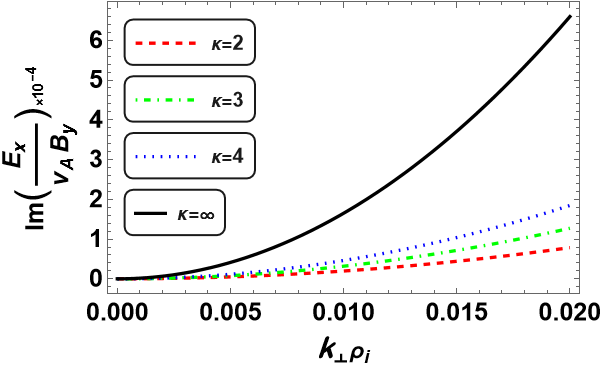}
                    }
        \caption{Normalized imaginary EM field $\text{Im} (\mathrm{E_x/v_{A}B_{y}})$ as a function of the normalized perpendicular wavenumber $k_{\perp}\rho_i$ for different values of $\kappa$. The parameters' values appropriate for the solar coronal region are; $\mathrm{v_{A}}\approx 1.85\times 10^{8}$ cm/sec, $\mathrm{v_{Ti}}\approx 1.9\times 10^{7}$ cm/sec, $\mathrm{v_{Te}} \approx 1.34\times 10^{9}$ cm/sec, $k_{\perp}/k_{\parallel}=(100 - 115)$ \citep{chen2012kinetic}, magnetic field \textbf{B} = $(50 - 100)$ G \citep{zirin1996mystery, gary2001plasma}, and temperature $\mathrm{T}>10^{6}$ Kelvin \citep{de2015recent}, respectively. In the left panel, the ratio of electron-to-ion temperature is $0.5$, and in the right panel, $\mathrm{T_e/T_i} \approx 0.4$. The black curve ($\kappa \rightarrow \infty$) represents the Maxwellian result.}
    \label{B}
\end{figure}

Numerically, we first assumed $\mathrm{T_e/T_i}=0.5$ \citep{chandran2010alfven} in the solar corona, as is shown in Fig. \ref{B} (left panel). We observed that the magnitude of the imaginary field ratio is enhanced for different values of $\kappa$
and the normalized perpendicular wavenumber $k_{\perp}\rho_i$. The plots indicate that the difference between the kappa and Maxwellian distributions becomes prominent at larger $k_{\perp}\rho_i$. 

Additionally, we evaluated the perturbed EM field at a constant $\mathrm{T_e/T_i} \approx 0.4$ \citep{mercier2015electron} to see how different temperatures affect the magnitude of $\text{Im} (\mathrm{E_x/v_{A}B_{y}})$. As is shown in the right panel, it is evident that the magnitude of $\text{Im} (\mathrm{E_x/v_{A}B_{y}})$ decreases when $\mathrm{T_e/T_i} = 0.4$. One possible reason for this behavior is that according to Eq. (\ref{eq:15}), the temperature is directly linked to the field ratio. Consequently, the magnitude of $\text{Im} (\mathrm{E_x/v_{A}B_{y}})$ decreases for smaller values of $\mathrm{T_e/T_i}$ and increases for larger values of $\mathrm{T_e/T_i}$.

\begin{figure}[h!]
        \centering
       {
            \includegraphics[width=.47\linewidth]{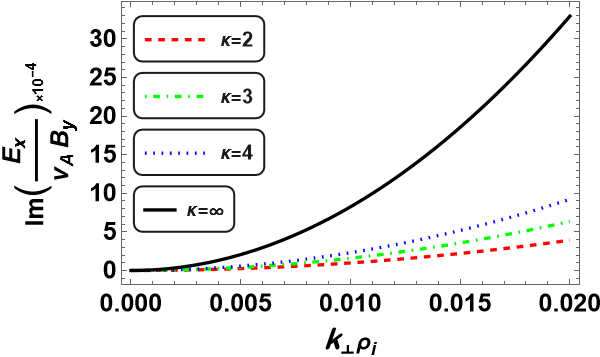}
                    }\quad
        {
            \includegraphics[width=.47\linewidth]{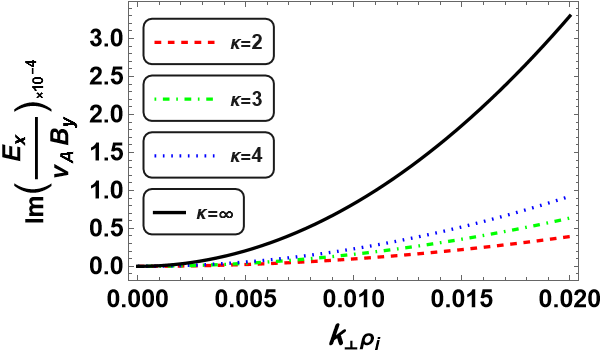}
                    }
        \caption{Normalized imaginary EM field $\text{Im} (\mathrm{E_x/v_{A}B_{y}})$ as a function of the normalized perpendicular wavenumber $k_{\perp}\rho_i$ for different values of $\kappa$. The parameters' values are the same as those in Fig. (\ref{B}). In the left and right panels, $\mathrm{T_e/T_i}=2$ and $\mathrm{T_e/T_i} \approx 0.2$, respectively.}
    \label{C}
\end{figure}

Moreover, \cite{barik2020nonresonant} studied the non-resonant instability of KAWs with $\kappa$-electrons using an electron-to-ion temperature ratio of $\mathrm{T_e/T_i}=2$. While their findings directly apply to the auroral region of Earth’s magnetosphere, they can also be extended to other magnetospheric regions, solar wind, and interplanetary medium, where ion beams, velocity shear, and non-Maxwellian electrons are present. We evaluated the normalized imaginary perturbed EM field $\text{Im} (\mathrm{E_x/v_{A}B_{y}})$ using this temperature ratio (Fig. \ref{C}, left panel) and found that the magnitude of  $\text{Im} (\mathrm{E_x/v_{A}B_{y}})$ is significantly increased and decreases when $\mathrm{T_e/T_i}=1/5$ (right panel). These evaluations confirm that the temperature ratio directly influences the magnitude of the perturbed EM field.

For clarity, we provide Table (Tab. \ref{tab}) with the selected parameters relevant to the solar coronal region. These values are representative of the solar corona and solar wind conditions. We used these parameters to discuss and develop our model, ensuring that the analysis aligns with the physical properties of these environments.
\begin{table}[h!]
  \begin{center}
    \caption{The Model Parameters}
    \label{tab}
    \begin{tabular}{l c c c c c c c c r}
      \hline
        \hline
    \textbf{Parameters} & & & & \textbf{Values}\\
      \hline
      $\textbf{B}$ &  & &  & 50 - 100 G\\
      $n$ &  & &  & $5 \times 10^{9}$ cm\textsuperscript{-3}\\
      $k_{\perp}/k_{\parallel}$ &  & &  & 100 - 115\\    
      h &  & &  & 0.05 - 0.1 R\textsubscript{Sun}\\
      $k_{\perp} \rho_i$ &  & &  & 0.01 - 0.1\\
      $\mathrm{T}$ &  & &  & $> 10^6$ K\\
      $\mathrm{v_A}$ &  & &  & $1.85 \times 10^{8}$ cm/sec\\
      $\mathrm{v_{Ti}}$ &  & &  & $1.9 \times 10^{7}$ cm/sec\\
      $\mathrm{v_{Te}}$ &  & &  & $1.34 \times 10^{9}$ cm/sec\\
      $a$ &  & &  & $\sim 7 \times 10^{7}$ cm\\
      $\mathrm{S_0}$ &  & &  & $10^{2}$ - $10^{4}$ W m\textsuperscript{-2}\\
      \hline
    \end{tabular}
  \end{center}
\end{table}

In the present study, we focus less on the real part of the perturbed EM field ratio, which is extensively discussed by \cite{Khan2019a,khan2020solar}. Our findings show that the imaginary part of the EM field, while relatively small in magnitude in the suprathermal environment, plays a crucial role in the solar coronal region. Previous studies, such as \cite{Lysak2003, khan2019distinct, lysak2023kinetic}, often neglected these minor contributions due to their focus beyond the solar coronal regions. However, we find that these subtle contributions are significant in the warm plasma of the solar corona. Drawing from \cite{Lysak2003}, we acknowledge that the small imaginary part of the $\text{Im} (\mathrm{E_x/v_{A}B_{y}})$ ratio may appear inconsequential near the Sun. In previous work \citep{Ayaz2024}, we evaluated this field ratio in a Cairns-distributed plasma and found that larger nonthermal parameters significantly influence the imaginary field ratio's magnitude. Our current observations align with our prior work, indicating that for smaller $\kappa$ values, the wave exhibits electrostatic characteristics consistent with \cite{lysak2023kinetic} findings. Additionally, our observations resonate with Polar observations by \cite{Wygant2002}, which provide empirical evidence of the ion gyroradius effect amplifying the $\mathrm{E_x/B_y}$ ratio.

\begin{figure}[h!]
        \centering
       {
            \includegraphics[width=.47\linewidth]{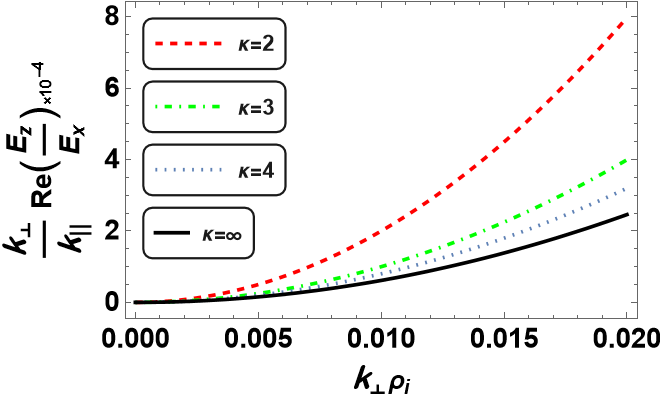}
                    }\quad
        {
            \includegraphics[width=.47\linewidth]{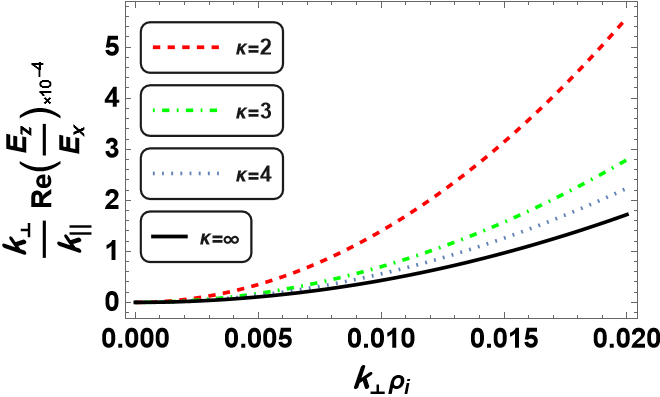}
                    }
        \caption{Normalized real EM field $\frac{k_{\perp}}{k_{\parallel}} \mathrm{Re} (\mathrm{E_z/E_x})$ as a function of normalized perpendicular wavenumber $k_{\perp}\rho_i$ for different values of $\kappa$. The parameters' values are the same as those in Fig. (\ref{B}). In the left and right panels, $\mathrm{T_e/T_i} \approx 0.5$ and $\mathrm{T_e/T_i} \approx 0.4$, respectively.}
    \label{D}
\end{figure}

The normalized real EM field ratio $\frac{k_{\perp}}{k_{\parallel}} \mathrm{Re} (\mathrm{E_z/E_x})$ is evaluated for different values of $\kappa$ and the electron-to-ion temperature ratio, as is illustrated in Fig. \ref{D}. Our analysis shows that $\frac{k_{\perp}}{k_{\parallel}} \mathrm{Re} (\mathrm{E_z/E_x})$ increases for smaller $\kappa$ values, with the difference becoming more pronounced at larger $k_{\perp}\rho_i$ values. For smaller $k_{\perp}\rho_i$, the difference between Maxwellian and kappa distributions diminishes. The suprathermal nature of the particles, characterized by the index $\kappa$, significantly influences the magnitude of $\frac{k_{\perp}}{k_{\parallel}} \mathrm{Re} (\mathrm{E_z/E_x})$.

We also examined the EM field for different temperature ratios. At a higher electron-to-ion temperature ratio of $0.5$ (left panel), the magnitude of the EM field is enhanced compared to lower temperature regions (right panel). Despite its relatively small magnitude, $\times 10^{-4}$, this part of $\frac{k_{\perp}}{k_{\parallel}} \mathrm{Re} (\mathrm{E_z/E_x})$ is typically neglected However, our study highlights the importance of this small contribution to the solar coronal regime.

\begin{figure}[h!]
        \centering
       {
            \includegraphics[width=.47\linewidth]{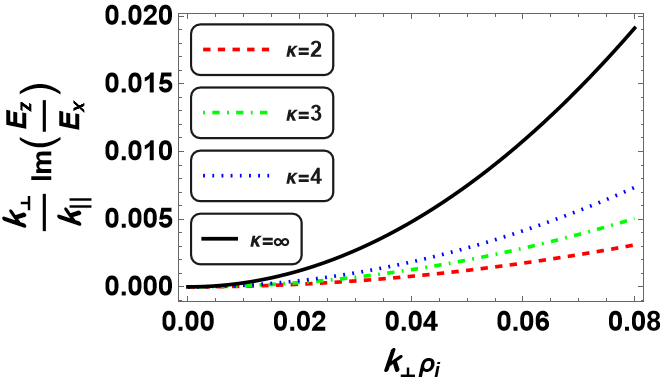}
                    }\quad
        {
            \includegraphics[width=.47\linewidth]{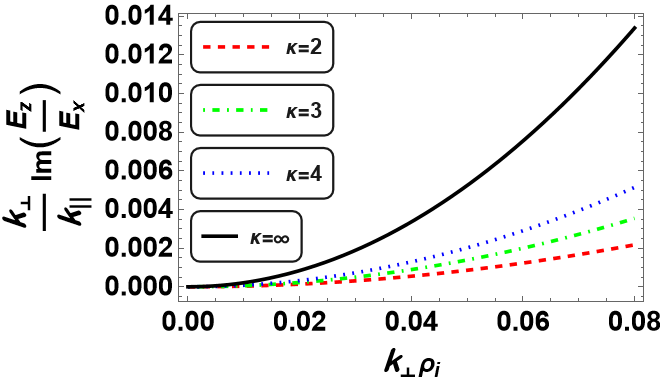}
                    }
        \caption{Normalized imaginary EM field $\frac{k_{\perp}}{k_{\parallel}} \mathrm{Im} (\mathrm{E_z/E_x})$ versus normalized perpendicular wavenumber $k_{\perp}\rho_i$ for different values of $\kappa$. The curves are generated using Eq. (\ref{eq:18}) with the parameters' values the same as those in Fig. (\ref{B}). In the left and right panels, $\mathrm{T_e/T_i}=0.5$ and $\mathrm{T_e/T_i} \approx 0.4$, respectively.}
    \label{E}
\end{figure}

Fig. \ref{E}, illustrates the normalized imaginary $\frac{k_{\perp}}{k_{\parallel}} \mathrm{Im} (\mathrm{E_z/E_x})$ for different values of $\kappa$ and $\mathrm{T_e/T_i}$. For a given $k_{\perp}\rho_i$, the magnitude of this ratio increases gradually for smaller $\kappa$ values. The difference between Maxwellian and kappa distributions becomes noticeable only at larger $k_{\perp}\rho_i$ values. Compared to the real part, $\frac{k_{\perp}}{k_{\parallel}} \mathrm{Re} (\mathrm{E_z/E_x})$ in Fig. \ref{C}, the imaginary part exhibits a similar trend with smaller $\kappa$ values, but the magnitude of $\frac{k_{\perp}}{k_{\parallel}} \mathrm{Re} (\mathrm{E_z/E_x})$ is approximately $\sim 10^3$ times smaller.

Furthermore, the trend of the curves for the imaginary part $\frac{k_{\perp}}{k_{\parallel}} \mathrm{Im} (\mathrm{E_z/E_x})$ is opposite to that of the real part (Fig. \ref{D}). Prior studies by \cite{Khan2019a} elaborate on these trends. While the detailed evaluation of $\frac{k_{\perp}}{k_{\parallel}} \mathrm{Im} (\mathrm{E_z/E_x})$ is extensively covered by \cite{khan2020solar}, we include these results here for consistency and comparison with our primary focus on $\mathrm{Re} (\mathrm{E_x/v_{A}B_y})$, which remains unexplored.

\subsection{The Poynting flux vectors of KAWs}
\begin{figure}[h!]
        \centering
       {
            \includegraphics[width=.47\linewidth]{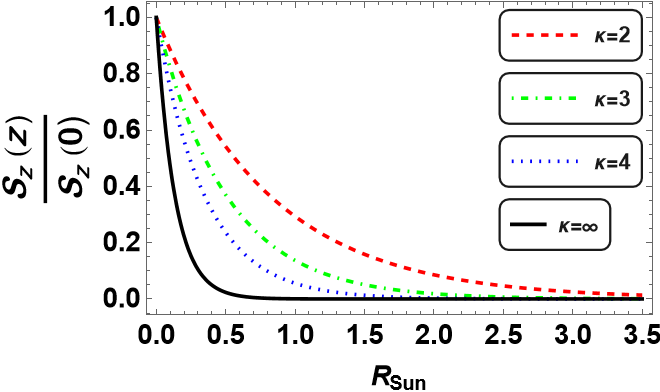}
                    }\quad
        {
            \includegraphics[width=.47\linewidth]{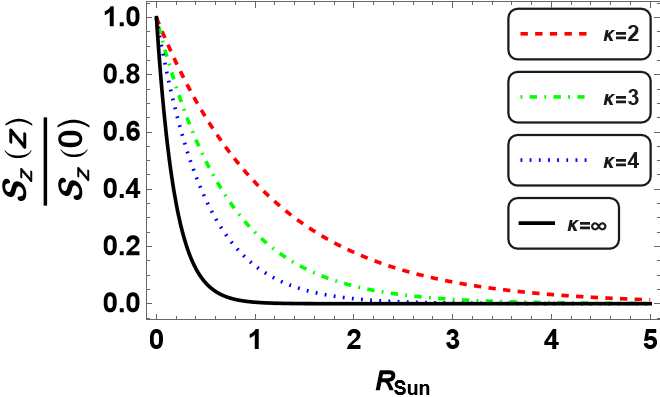}
                    }
        \caption{Normalized Poynting flux $\mathrm{S_{z}(z)/S_{z}(0)}$ as a function of the radius of the Sun (R\textsubscript{Sun}) for different values of $\kappa$. We assumed $k_{\perp}\rho_i \approx0.02$, h = 0.05 R\textsubscript{Sun}, and the other parameter values are the same ones that we used in Fig. \ref{B}. In the left panel, $\mathrm{T_e/T_i}=0.5$ and in the right panel it is $0.4$.}
    \label{F}
\end{figure}

\begin{figure}[h!]
        \centering
       {
            \includegraphics[width=.47\linewidth]{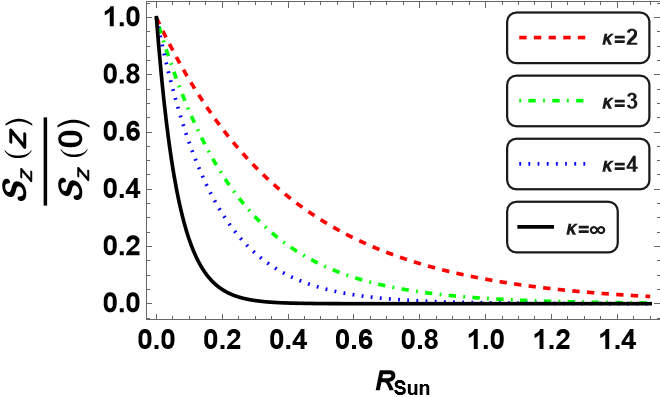}
                    }\quad
        {
            \includegraphics[width=.47\linewidth]{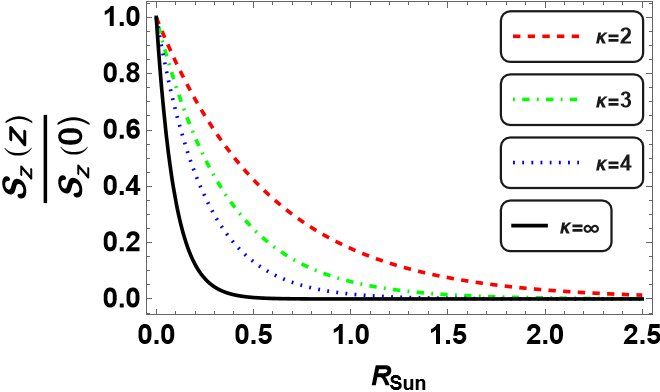}
                    }
        \caption{Normalized Poynting flux $\mathrm{S_{z}(z)/S_{z}(0)}$ as a function of the radius of the Sun R\textsubscript{Sun} for different values of $\kappa$ at fixed h = 0.1 R\textsubscript{Sun}. The parameter values are the same as those we used in Fig. (\ref{F}) with the left panel, $\mathrm{T_e/T_i}=0.5$, and $0.4$ in the right panel.}
    \label{G}
\end{figure}

Since the Poynting flux vector is key to determining how much EM energy the waves transfer to the plasma as they propagate from $z=0$, Figs. (\ref{F}) and (\ref{G}) show the Poynting flux vector of KAWs for various values of the index parameter $\kappa$, electron-to-ion temperature ratios $\mathrm{T_e/T_i}$, and heights h relative to the Sun's radius (R\textsubscript{Sun}). When $\mathrm{T_e/T_i}=0.5$ (Fig. \ref{F}, left panel), the Poynting flux rate $\mathrm{S_{z}(z)/S_{z}(0)}$ decreases moderately slower for smaller $\kappa$ values. For larger $\kappa$ values ($\kappa \rightarrow \infty$), representing the Maxwellian case, the flux rate decays rapidly, causing the wave to damp quickly. Conversely, with a temperature ratio of $0.4$ (Fig. \ref{F}, right panel), the wave damps at a slower rate, allowing it to transport energy over a more extended distance compared to when $\mathrm{T_e/T_i}=0.5$ with the same $\kappa$ variation. This variation in $\mathrm{S_{z}(z)/S_{z}(0)}$ for different $\kappa$ values has interesting implications. Smaller $\kappa$ values, indicating the presence of suprathermal particles, enable the wave to transport energy over longer distances. When the temperature ratio is lower, the wave transports energy over greater distances compared to higher temperature ratios. This suggests that at higher temperatures, a larger number of particles resonate, causing the Poynting flux to decay quickly. In recent work by \cite{khan2020solar}, this phenomenon is well elaborated for both isotropic and anisotropic particle temperatures, with analytic results consistent with our study.

Consider now a case in which we increased height h = 0.1 R\textsubscript{Sun} and evaluate $\mathrm{S_{z}(z)/S_{z}(0)}$ for different $\kappa$ values. The results show that $\mathrm{S_{z}(z)/S_{z}(0)}$ decreases rapidly for larger $\kappa$ values, while for smaller $\kappa$ values, the Poynting flux decays more gradually, as is depicted in Fig. \ref{G}. Different temperature ratios correspond to different spatial distances over which the wave transports energy. Specifically, with lower $\mathrm{T_e/T_i}$ ratios (right panel), the wave transports energy over a greater distance compared to higher $\mathrm{T_e/T_i}$ ratios. The flux rate also decays rapidly when h = 0.1 R\textsubscript{Sun}.

This analysis highlights the significant differences between Maxwellian and non-Maxwellian plasmas within the small distance range of R\textsubscript{Sun}, approximately $(0.4 - 5)$ R\textsubscript{Sun}. According to the Poynting flux expression (\ref{eq:26}), during energy transport, the wave converts its EM energy to Landau resonant electrons, with the conversion dependent on the stationary state described by the index $\kappa$. In kappa-distributed plasma, compared to Maxwellian plasma, the wave transfers its energy to resonant electrons over larger distances. This means that in the small-$\kappa$ regime, energy is distributed among a smaller number of suprathermal particles, making small $\kappa$ values beneficial for heating and/or accelerating plasma particles over long distances. This behavior may be observed in the solar wind and corona, where the Polar spacecraft has reported $\kappa \leq 10$ \citep{kletzing2003auroral} that fit the data well compared to Maxwellian distribution.

\begin{figure}[h!]
        \centering
       {
            \includegraphics[width=.47\linewidth]{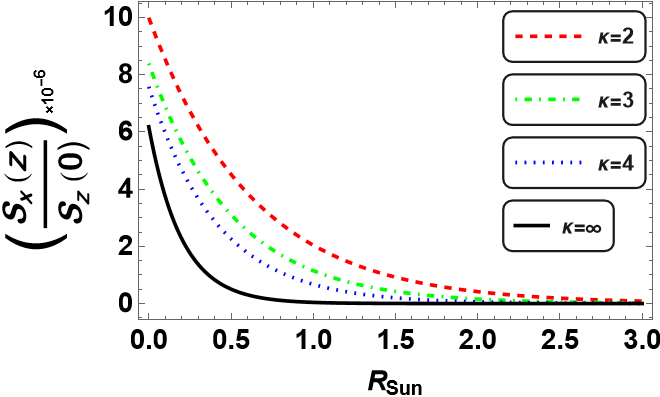}
                    }\quad
        {
            \includegraphics[width=.47\linewidth]{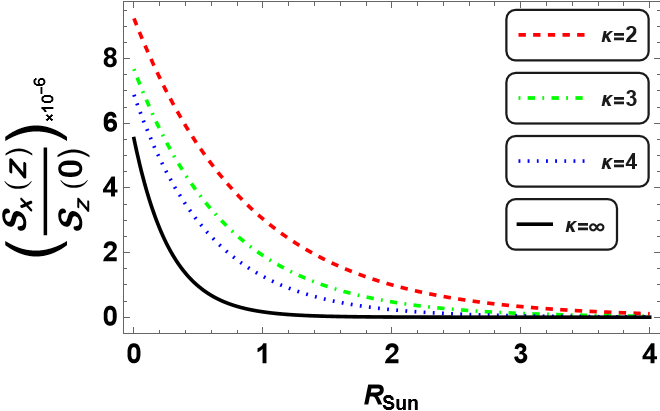}
                    }
        \caption{Normalized perpendicular Poynting flux $\mathrm{S_{x}(z)/S_{z}(0)}$ versus R\textsubscript{Sun} for different values of $\kappa$ at a fixed height of h = 0.05 R\textsubscript{Sun}. The parameter values are the same as those we used in Fig. (\ref{F}) with the left panel, $\mathrm{T_e/T_i}=0.5$, and $0.4$ in the right panel.}
    \label{H}
\end{figure}

\begin{figure}[h!]
        \centering
       {
            \includegraphics[width=.47\linewidth]{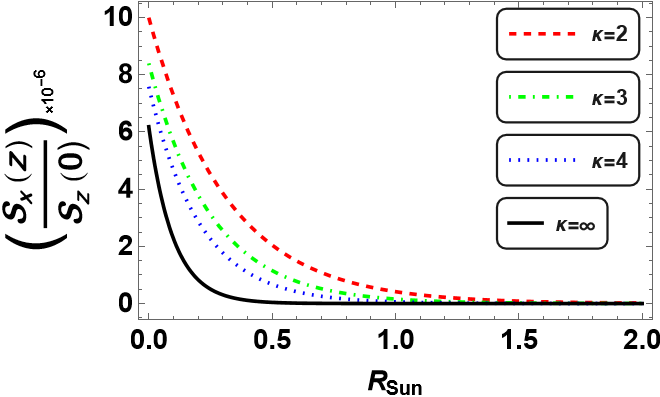}
                    }\quad
        {
            \includegraphics[width=.47\linewidth]{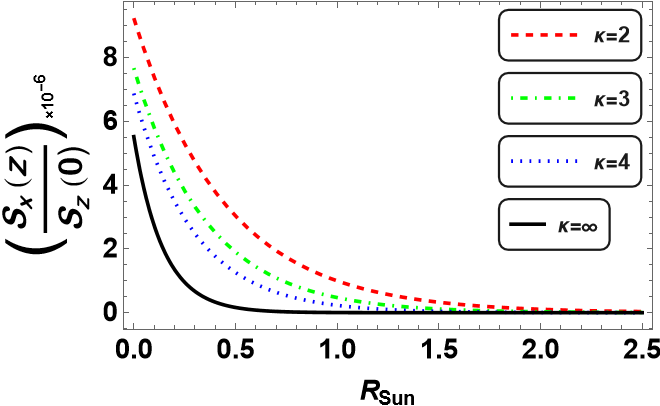}
                    }
        \caption{Normalized perpendicular Poynting flux $\mathrm{S_{x}(z)/S_{z}(0)}$ versus R\textsubscript{Sun} for different values of $\kappa$ at a fixed height of h = 0.1 R\textsubscript{Sun}. The parameter values are the same as those we used in Fig. \ref{F} with the left panel, $\mathrm{T_e/T_i}=0.5$, and $0.4$ in the right panel.}
    \label{I}
\end{figure}

The preceding analysis of $\mathrm{S_{z}(z)/S_{z}(0)}$ provides insights into the parallel energy transportation of KAWs. While this parallel Poynting flux and energy delivery by KAWs has been extensively studied (\cite{Lysak2003,khan2019distinct,khan2020solar,lysak2023kinetic}, and references therein) consider now the perpendicular energy transport of KAWs. Given that $k_{\parallel}\ll k_{\perp}$, KAWs typically carry less energy in the perpendicular direction. Figs. (\ref{H} - \ref{I}) illustrate how KAWs transport energy perpendicularly. The magnitude of the perpendicular Poynting flux $\mathrm{S_{x}(z)/S_{z}(0)}$ decreases for larger $\kappa$ values and increases for smaller $\kappa$ values. In the left panel, with a temperature ratio of $\mathrm{T_e/T_i} = 0.5$, $\mathrm{S_{x}(z)/S_{z}(0)}$ decays faster compared to the right panel, where $\mathrm{T_e/T_i} = 0.4$.

For the case of h = 0.1 R\textsubscript{Sun}, the magnitude of $\mathrm{S_{x}(z)/S_{z}(0)}$ follows the same trend observed previously when 
h = 0.05 R\textsubscript{Sun}, for the same variations in $\kappa$ and $\mathrm{T_e/T_i}$, as is shown in Fig. \ref{I}. The primary difference lies in the distance (R\textsubscript{Sun}) over which the wave transports energy. At h = 0.05 R\textsubscript{Sun}, the wave carries energy over a longer distance (Fig. \ref{H}) compared to h = 0.1 R\textsubscript{Sun} (Fig. \ref{I}). This difference in the spatial extent of energy transport is due to the contribution of the electric field in the x direction, as is described in Eq. (\ref{eq:27}). This electric field, combined with the strong magnetic field, significantly influences the energy transport of KAWs in the perpendicular direction.

\subsection{The net power delivered by KAWs}
The Poynting flux expressions for parallel and perpendicular directions ($\mathrm{S_x}$ and $\mathrm{S_z}$) are used to determine the total power transfer rate of KAWs in a solar flux tube loop (Fig. \ref{A}, right panel); that is, the total power transfer rate ($\mathrm{I_x}/\mathrm{I_z}$). The flux tube is characterized by its height, $\mathrm{h}$, and radius or width $a$. The height - represented by the parallel Poynting flux vector $\mathrm{S_z}$, indicates that more energy is transported in the parallel direction by the KAWs. Conversely, the width of the loop, characterized by the perpendicular Poynting flux $\mathrm{S_x}$, shows that less energy is transported in the perpendicular direction. We need both $S_x$ and $S_z$ to determine the total power delivered within the flux tube loop.

\begin{figure}[h!]
        \centering
       {
            \includegraphics[width=.47\linewidth]{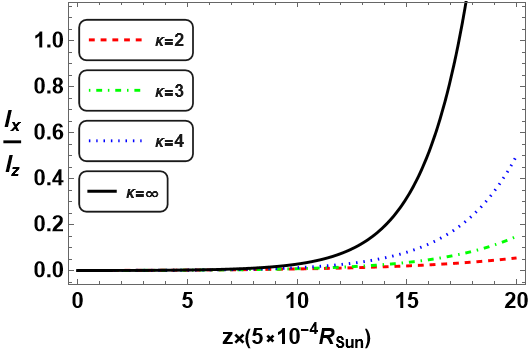}
                    }\quad
        {
            \includegraphics[width=.47\linewidth]{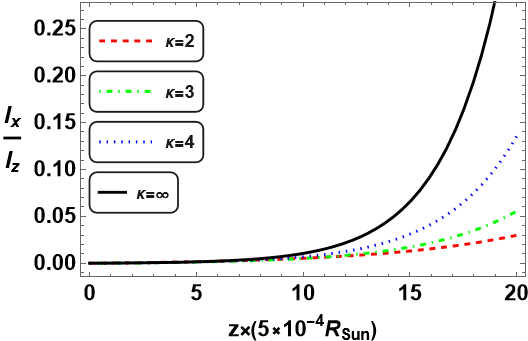}
                    }
        \caption{The normalized power transfer rate ($\mathrm{I_x}/\mathrm{I_z}$) as a function of the normalized distance $z \times 0.0005$ $R_{\text{Sun}}$, at a fixed height h = 0.1 R\textsubscript{Sun}. The radius of the circular crosssection, $a$, is assumed to be $7 \times 10^7$ cm, with other parameters consistent with those used in Fig. \ref{F}. In the left panel, $\mathrm{T_e/T_i} = 0.5$, and in the right panel, $\mathrm{T_e/T_i} = 0.4$. The power transfer rate is significantly influenced by the suprathermal particles, characterized by $\kappa$.}
    \label{J}
\end{figure}

Fig. \ref{J} shows that the total power transfer rate ($\mathrm{I_x}/\mathrm{I_z}$) is significantly influenced by different values of $\kappa$. For larger $\kappa$ values ($\sim$ Maxwellian conditions), $\mathrm{I_x}/\mathrm{I_z}$  increases sharply over a shorter distance, indicating that the wave delivers its power to the particles over a small spatial extent. For smaller $\kappa$ values, indicative of more suprathermal particles, the power transfer rate increases more gradually over larger distances, meaning the wave transports power over a longer spatial extent. This occurs because the suprathermal environment facilitates more particle contributions to energy transport, allowing the wave to carry power over greater distances.

We also evaluated $\mathrm{I_x}/\mathrm{I_z}$ for different temperature ratios. Assuming $\mathrm{T_e/T_i}=0.5$ (left panel), the power transfer rate is lower compared to the case of $\mathrm{T_e/T_i}=0.4$ (right panel). At higher temperatures, the wave transports more power over a longer distance. When the temperature is lower, the wave transports less power over a shorter distance.

\begin{figure}[h!]
        \centering
       {
            \includegraphics[width=.47\linewidth]{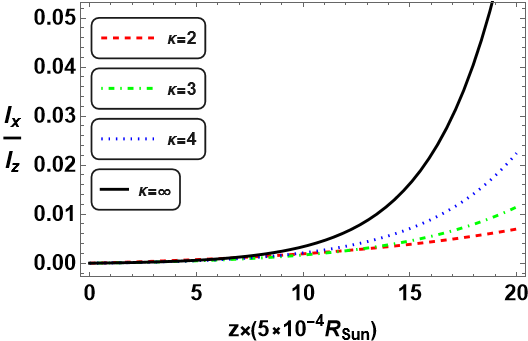}
            \label{a}
        }\quad
        {
            \includegraphics[width=.47\linewidth]{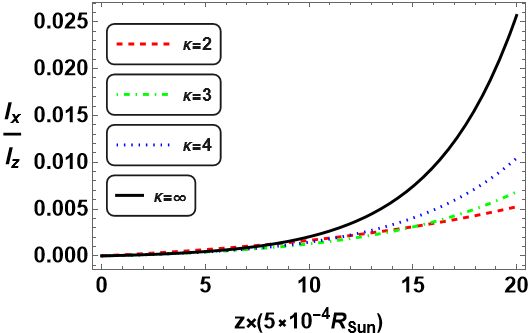}
            \label{b}
        }
        \caption{The normalized $\mathrm{I_x}/\mathrm{I_z}$ versus distance $z\times 0.0005$ R\textsubscript{Sun} at fixed h = 0.05 R\textsubscript{Sun}. The parameters are the same as in Fig. \ref{J}. In the left panel, $\mathrm{T_e/T_i}=0.5$, and $0.4$ in the right panel.}
    \label{K}
\end{figure}

With h = 0.05 R\textsubscript{Sun} and the same values for $\kappa$ and the electron-to-ion temperature ratios, the normalized power transfer rate increases more rapidly, as is shown in Fig. \ref{K}. The power transfer rate accelerates more quickly in regimes with larger $\kappa$ values, while it progresses more slowly in regimes with smaller $\kappa$ values. We also observe that the power transfer rate in the h = 0.05 R\textsubscript{Sun} case is faster compared to the h = 0.1 R\textsubscript{Sun} scenario.

In \cite{Ayaz2024}, we evaluated the power delivery rate of KAWs in a Cairns-distributed plasma and found that the power transfer rate is enhanced in the presence of a larger number of nonthermal particles. Our findings here are consistent with that, reaffirming that the power transfer rate is significantly influenced by suprathermal particles. Furthermore, our examination of the different heights implies that in regions closer to the Sun, where h is smaller, KAWs can transport energy more efficiently over shorter distances, especially in plasmas with higher suprathermal particle populations.

\begin{figure}[h!]
        \centering
        {
            \includegraphics[width=.47\linewidth]{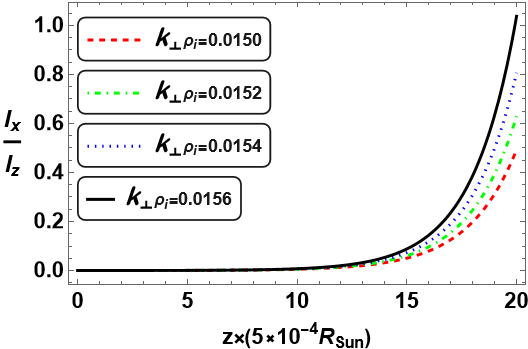}
            \label{c}
        }\quad
        {
            \includegraphics[width=.47\linewidth]{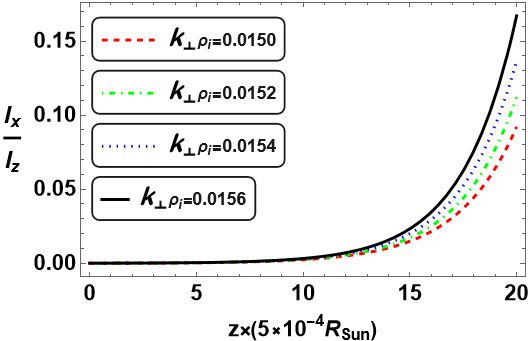}
        }
  \caption{The power transfer rate ($\mathrm{I_x}/\mathrm{I_z}$) as a function of $z\times 0.0005$ R\textsubscript{Sun} at a fixed h = 0.05 R\textsubscript{Sun} for different values of $k_{\perp}\rho_i$. The parameters are the same as in Fig. \ref{J}. In both panels, we assumed h = 0.05 R\textsubscript{Sun} and larger values of $\kappa$ with $\mathrm{T_e/T_i}=0.5$ (left), and $\mathrm{T_e/T_i}=0.4$ (right panel).}
  \label{L}
\end{figure}

Besides variation in $\kappa$, we also evaluated the power transfer rate for different values of the normalized wavenumber $k_{\perp}\rho_i$, as is shown in Fig. \ref{L}. For simplicity, we assumed larger values of $\kappa$ -- the Maxwellian situation -- and observed that $\mathrm{I_x}/\mathrm{I_z}$ is significantly influenced by even minor changes in $k_{\perp}\rho_i$. This indicates that small changes in $k_{\perp}\rho_i$ lead to an increased power delivery rate. Specifically, for larger $k_{\perp}\rho_i$ values, the wave transports power more rapidly (the black curve). Conversely, for smaller $k_{\perp}\rho_i$ values, the wave transport power is moderately weaker (the red curve).

Additionally, when we decreased the temperature ratio to 0.4 (right panel), the power transfer rate became sharper, and the wave transported power at an even faster rate. Physically, this implies that the efficiency of power transfer by KAWs is highly sensitive to the wavenumber and temperature ratios. In regions with higher $k_{\perp}\rho_i$ or lower temperature ratios, KAWs can deliver power more effectively over shorter distances. The results align with our previous study \citep{Ayaz2024}, reinforcing the significance of nonthermal particles in enhancing the power transfer rate of KAWs.

\subsection{Net resonance velocity of the particles}
\begin{figure}[h!]
        \centering
        {
            \includegraphics[width=.47\linewidth]{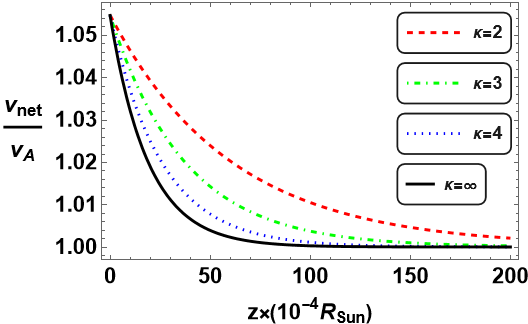}
            \label{e}
        }\quad
        {
            \includegraphics[width=.47\linewidth]{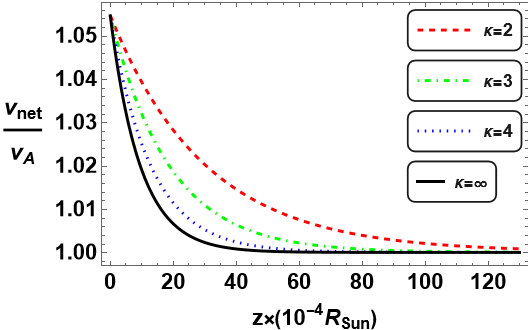}
            \label{f}
            }
  \caption{Normalized parallel resonance velocity $\mathrm{v_{res\perp}/v_{A}}$ as a function of $z(\times 10^{-4})$ R\textsubscript{Sun} for different values of $\kappa$ at fixed $\mathrm{T_e/T_i}=0.5$. We assumed $\mathrm{S(0)} = 10^2 - 10^4$ W cm\textsuperscript{-2} \citep{Srivastava2017}, density $n_0\approx 5\times 10^{9}$ cm\textsuperscript{-3} \citep{morton2015investigating}, $k_{\perp}\rho_i\approx 0.02$, and the other parameters are the same as those in Fig. \ref{L}. In the left panel, h = 0.05 R\textsubscript{Sun} and h = 0.1 R\textsubscript{Sun} in the right panel.}
  \label{M}
\end{figure}

Fig. \ref{M}, illustrates the normalized resonant velocity, $\mathrm{v_{res}/v_{A}}$, for different values of $\kappa$ and height h. The particle velocity, $\mathrm{v_{res}/v_{A}}$, is significantly influenced by $\kappa$ and decays at a moderately slower rate for smaller $\kappa$ values. This provides insights into wave-particle interactions, showing that particles accelerate to higher speeds, enabling energy transport and heating over longer distances.

We also evaluated $\mathrm{v_{res}/v_{A}}$ for both high and low temperatures and found no significant effect of temperature on the resonant velocity. The normalized $\mathrm{v_{res}/v_{A}}$ represents the heating and acceleration of charged particles in the parallel direction through Landau resonance. This process is described in expression (\ref{eq:35}), with corresponding plots shown in Fig. \ref{M}. We further investigated $\mathrm{v_{res}/v_{A}}$ for different heights h. At h = 0.05 R\textsubscript{Sun} (left panel), particles accelerate and transport energy over longer distances compared to h = 0.1 R\textsubscript{Sun} (right panel). For h = 0.1 R\textsubscript{Sun}, particles heat the plasma over shorter distances (i.e., $\sim 100$). Nearby (0.05 R\textsubscript{Sun}), particles accelerate over longer distances because more suprathermal particles contribute to wave-particle interactions, and facilitate extended heating distances. This detailed analysis gives the significant role of $\kappa$ and height in the dynamics of energy transport and particle acceleration by KAWs in the solar corona.

\begin{figure}[h]
        \centering
        {
            \includegraphics[width=.47\linewidth]{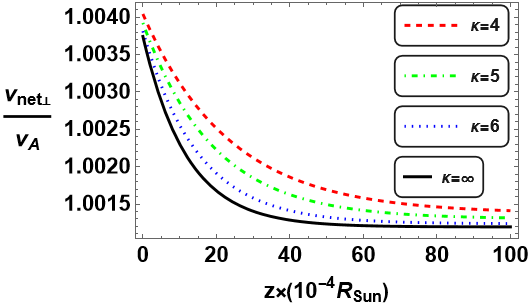}
            \label{g}
        }\quad
        {
            \includegraphics[width=.47\linewidth]{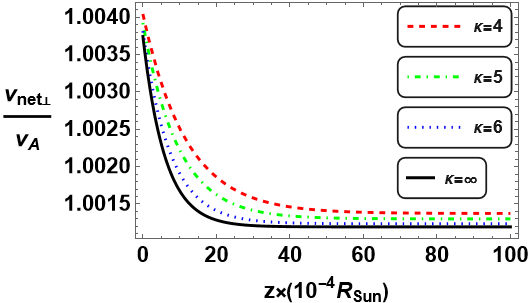}
            \label{h}
}
  \caption{Normalized perpendicular resonance velocity, $\mathrm{v_{res\perp}/v_{A}}$, versus normalized distance $z(\times 10^{-4})$ R\textsubscript{Sun} for different $\kappa$ values at a fixed $\mathrm{T_e/T_i}=0.5$. The parameters are consistent with those in Fig. \ref{M}. In the left panel, h = 0.05 R\textsubscript{Sun} and the right panel represents h = 0.1 R\textsubscript{Sun}.}
  \label{N}
\end{figure}

Fig. \ref{N} shows the normalized perpendicular net velocity $\mathrm{v_{res \perp}/v_{A}}$ of KAWs for different values of $\kappa$. KAWs carry less energy in perpendicular directions. We observe that $\mathrm{v_{res \perp}/v_{A}}$ decreases at a slower rate for small $\kappa$ values and faster for larger $\kappa$ values. In the case of smaller $\kappa$ values, more suprathermal particles participate in wave-particle interactions, significantly affecting $\mathrm{v_{res \perp}/v_{A}}$ and accelerating and heating the plasma over larger distances. For larger $\kappa$ values, the net resonant speed of particles decreases more rapidly, resulting in shorter acceleration distances.

We also evaluated $\mathrm{v_{res \perp}/v_{A}}$ for different heights, $\mathrm{h}$. When h = 0.05 R\textsubscript{Sun}, particles heat the plasma over extended distances (left panel). For h = 0.1 R\textsubscript{Sun}, the resonant speed of particles decays very quickly, limiting the acceleration distance (right panel). Furthermore, we assessed $\mathrm{v_{res \perp}/v_{A}}$ for different electron-to-ion temperature ratios and found no significant effect on the normalized perpendicular resonant speed. These variations in $\mathrm{v_{res \perp}/v_{A}}$ for different heights and suprathermal particles, characterized by the index $\kappa$, illustrate how resonant particles can heat and accelerate the plasma over extended distances.

Very recently, \cite{rivera2024situ} investigated the heating and acceleration of solar wind plasma as it moves from the outer edge of the solar corona to the inner heliosphere, supported by observations from the PSP and Solar Orbiter. The study shows that large-amplitude Alfv\'{e}n waves contribute significantly to both plasma heating and acceleration. This observation can be linked to our study of KAWs, which play a central role in transferring energy through wave-particle interactions, heating the plasma, and accelerating charged particles in the solar wind and solar corona. The analytical results are compared with Rivera et al., who found the Alfv\'{e}n speed was approximately $\mathrm{v_A} \approx 4.35 \times 10^7$ cm/s at a distance of 13.5 R\textsubscript{Sun}. To ensure consistency with this value, we assumed a magnetic field strength of \textbf{B} = 10 G in our analytical approach. Our calculations yielded an Alfv\'{e}n speed of about $\mathrm{v_A} \approx 3 \times 10^8$ cm/s, which is largely consistent with Rivera et al.'s results, with the minor discrepancy may likely be due to the difference in distance, as we considered 0.05 to 0.1 R\textsubscript{Sun}.

Furthermore, Rivera et al. reported a bulk particle speed (i.e., proton's speed) of roughly $\mathrm{v_{\alpha}} \approx 3.96 \times 10^7$ cm/s at the same 13.5 R\textsubscript{Sun} region. However, our analytical results, particularly for the resonant particle speed, indicate that particles reach a net resonant speed around $\sim 3.15 \times 10^8$ cm/s at 0.05 to 0.1 R\textsubscript{Sun}. The findings of our research are typically consistent with the observational analysis, especially the net parallel and perpendicular resonance speed of the particles contributing to acceleration and heating processes in the solar wind and solar corona regions. 

\subsection{Group velocity and the damping length of KAWs}
\begin{figure}[h!]
    \centering
    \includegraphics[width=0.5\linewidth]{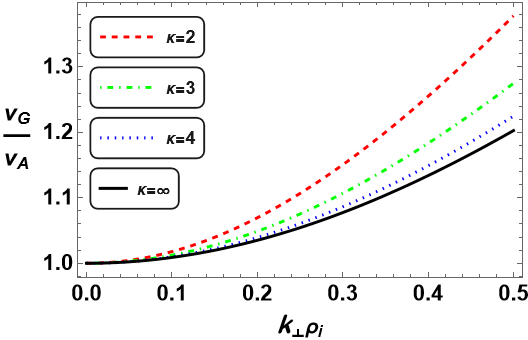}
    \caption{Normalized group velocity ($\mathrm{v_G/v_{A}}$) as a function of the normalized wavenumber $k_{\perp}\rho_i$ for different values of $\kappa$. We assumed $\mathrm{T_e/T_i}=0.5$ and the other parameter values are the same that we used in the previous figures. The normalized group velocity is enhanced for smaller values of $\kappa$. }
    \label{O}
\end{figure}

The analysis of group speed is also very important; it tells us how the energy flows. The normalized group velocity ($\mathrm{v_G/v_{A}}$) of KAWs is evaluated for different values of the suprathermal parameter $\kappa$ as is shown in Fig. \ref{O}. The results indicate a significant $\mathrm{v_G/v_{A}}$ enhancement for smaller $\kappa$ values,  in contrast to the Maxwellian case ($\kappa \rightarrow \infty$). Lower $\kappa$ values, and hence greater presence of suprathermal particles, collectively increase the group velocity of the wave.

\begin{figure}[h!]
        \centering
        {
            \includegraphics[width=.47\linewidth]{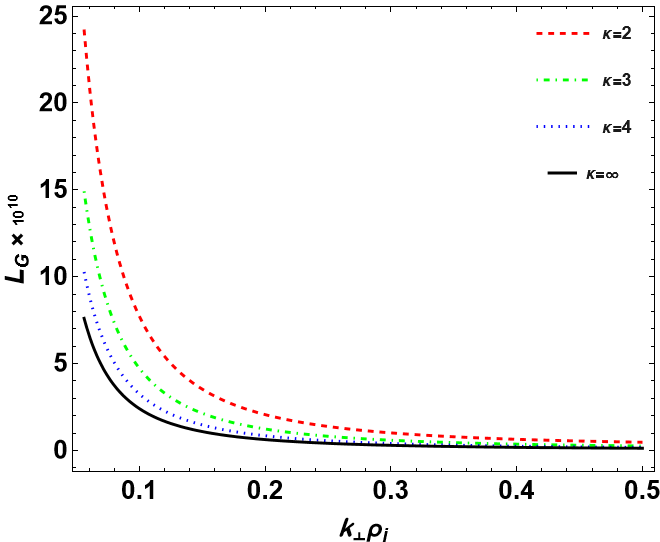}
        }\quad
        {
            \includegraphics[width=.47\linewidth]{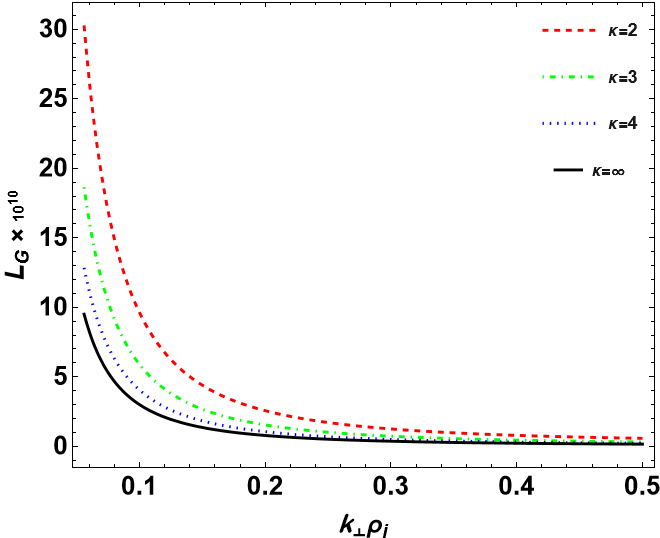}
        }
        \caption{The damping length ($\mathrm{L_G}$) as a function of the normalized perpendicular wavenumber $k_{\perp}\rho_{i}$ for different values of $\kappa$. The graphs are plotted using Eq. (\ref{eq:40}) with the same parameter values as those which we assumed in Fig. \ref{A}. In the left panel, $\mathrm{T_e/T_i}=0.4$, and in the right panel, $\mathrm{T_e/T_i}=0.5$. The damping length is significantly influenced by different values of electron-to-ion temperature ratios and $\kappa$.}
    \label{P}
\end{figure}

Finally, we evaluate the characteristic damping length ($\mathrm{L_G}$) - provides insight into how far KAWs and particles can transport energy before being damped, for different values of $\kappa$, shown in Fig. \ref{P}. $\mathrm{L_G}$ decreases for larger $\kappa$ values and increases for smaller $\kappa$ values. Waves (KAWs) damp quickly in the larger $\kappa$ regimes and at a moderately slower rate when $\kappa$ values are smaller. This implies that smaller $\kappa$ values are advantageous for heating or accelerating plasma particles over long distances, which may be observed in the solar corona and solar wind regions.
 
We also investigated $\mathrm{L_G}$ under different temperature conditions. In the right panel, when the electron-to-ion temperature ratio is 0.4, the magnitude of $\mathrm{L_G}$ increased for the same variation in $\kappa$. This indicates that lower temperature values increase the damping length, allowing particles to transport energy for longer distances before damping. Conversely, in the higher temperature scenario ($\mathrm{T_e/T_i}=0.5$, left panel), the magnitude of $\mathrm{L_G}$ is reduced; that is, higher temperatures cause KAWs to decay faster and heat the plasma over shorter distances. This analysis provides a good approximation and estimate of how KAWs get damped and how far they can transport energy before damping.

\begin{figure}[h!]
        \centering
        {
            \includegraphics[width=.47\linewidth]{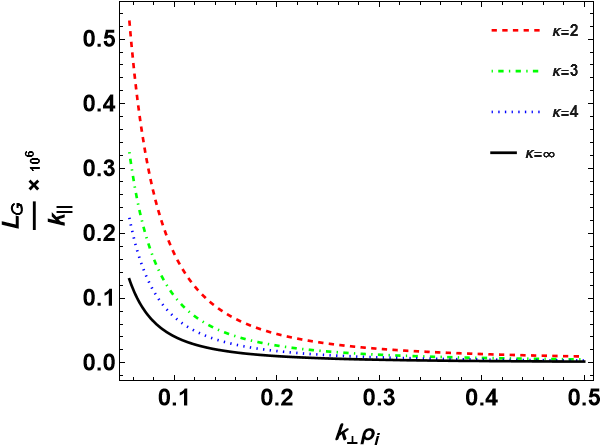}
        }\quad
        {
            \includegraphics[width=.47\linewidth]{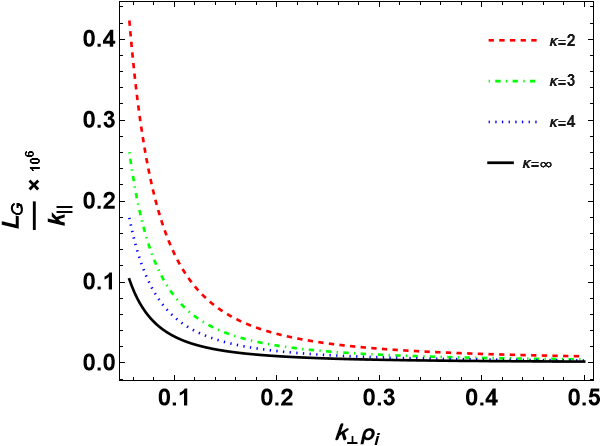}
        }
        \caption{Damping length ($\mathrm{L_G}$) and the parallel wavevector ($k_{\parallel}$) as a function of the normalized perpendicular wavenumber $k_{\perp}\rho_{i}$ for different values of $\kappa$. The parameter values are the same as the ones that we assumed in Fig. \ref{P}. In the left panel, $\mathrm{T_e/T_i}=0.4$, and in the right panel, $\mathrm{T_e/T_i}=0.5$.}
    \label{Q}
\end{figure}

The ratio of $\mathrm{L_G}$ and the parallel wavevector ($k_{\parallel}$) as a function of $k_{\perp}\rho_{i}$ is evaluated for different values of $\kappa$ (see Fig. (\ref{Q})). According to our chosen parameters and the required conditions of KAWs, we have $k_{\parallel}\ll k_{\perp}$. In this work, we have $k_{\perp}/k_{\parallel} \approx$ (100 - 115), which corresponds to an angle of $\theta \approx$ (89.5 - 90). Therefore, it is recommended to have the $k_{\parallel}$ 100 times smaller than $k_{\perp}$. Based on our model parameters, we have $k_{\perp}\rho_i \approx 0.01$, which is consistent with \cite{singh2019anisotropic} who studied KAWs heating in the solar corona. Thus, in our model, $k_{\parallel} \approx \times 10^{-6}$ cm\textsuperscript{-1}. Substituting the parameter values in Eq. (\ref{eq:40}), we get $\mathrm{L_G} \approx \times 10^{10}$ - equivalent to R\textsubscript{Sun}, which is shown in Fig. (\ref{P}). The ratio of $\mathrm{L_G}$ and $k_{\parallel}$ gives $\sim \times 10^4$. Based on this analysis, the magnitude of $\mathrm{L_G}/k_{\parallel}$ is $10^4$ times smaller than $\mathrm{L_G}$ alone, as is illustrated in Fig. (\ref{Q}). 
This implies that KAWs can propagate over considerable distances with minimal damping. This efficiency in energy transport suggests that KAWs play a vital role in transferring energy (for example, released during magnetic reconnection) throughout the solar corona, enhancing particle heating and acceleration while highlighting the dominance of perpendicular structures in their dynamics.

\section{Discussion and conclusion}
We present a detailed study of the energy transportation and heating phenomena of KAWs in the solar corona using kappa-distributed plasmas. Our focus is on the perturbed EM field ratios for different values of $\kappa$ and the effects of electron-to-ion temperature ratios. Specifically, we investigate the imaginary field ratios ($\text{Im} (\mathrm{E_x/v_{A} B_y})$ and $k_{\perp}/k_{\parallel} \text{Im} (E_z/E_x)$), which yields interesting results. For regions with larger wavenumbers and smaller suprathermal particle populations, the magnitude of the imaginary fields is slightly enhanced. This provides valuable insights into how EM perturbations are influenced by suprathermal particles ($\kappa$) and the electron-to-ion temperature.

We observe that the perturbed field ratios are significantly enhanced at higher temperatures. Our primary interest lies in determining how the energy stored in these EM fields is converted into the heating and acceleration of charged particles. By utilizing these field ratios in the Poynting flux expression, we gain knowledge about how KAWs transport energy to particles. Our analytical results align with previous studies by \cite{Lysak2003}, \cite{khan2020solar}, and \cite{lysak2023kinetic}. Additionally, our findings are consistent with observational data from \cite{Wygant2002} and the numerical investigation in our recent work \citep{Ayaz2024}. We found that the perturbations in the imaginary field ratios are significantly affected by the larger nonthermal parameters. This comprehensive analysis advances our understanding of the role of KAWs in the solar corona's energy dynamics.

Understanding the perturbed EM fields is important for confirming wave characteristics and determining how energy is stored and transported by KAWs in the coronal region \citep{khan2020solar}. This study focuses on the Poynting flux vector, which provides essential information about how KAWs carry energy along ambient magnetic field lines. Using an analytical approach (i.e., Eq. \ref{eq:26}), we demonstrate that KAWs transport significant energy parallel to these field lines. Our findings indicate that the Poynting flux rate decays more slowly for smaller values of $\kappa$. Physically, this suggests that an abundance of suprathermal particles contributes to slower energy decay, allowing the wave to transport energy over longer distances. The examination of the parallel Poynting vector ($\mathrm{S_{z}(z)/S_{z}(0)}$) shows that KAWs transport more energy in that along the magnetic field lines. We also evaluated ($\mathrm{S_{z}(z)/S_{z}(0)}$) by varying height and temperature. When h = 0.1 R\textsubscript{Sun}, which is farther from the Sun, the wave's energy transport decays more rapidly compared to h = 0.5 R\textsubscript{Sun}, a closer distance. This behavior is expected, as waves transport energy over longer spatial extents in nearer regions and decay more quickly in farther regions.

Additionally, we evaluated the effect of different temperature ratios on the energy transportation of KAWs. Higher temperatures result in longer energy transport distances and slower decay rates, while lower temperatures lead to faster decay. This analysis aligns with previous studies by \cite{Khan2019a,khan2019distinct,khan2020solar} that found that KAWs transport more energy when suprathermal particles are abundant. Our recent work \citep{Ayaz2024} also supports these findings, reinforcing the significance of suprathermal particles in the energy dynamics of KAWs in the solar corona.

\cite{zank2018theory} adopted the perspective of \cite{markovskii2006dissipation} and studied perpendicular and parallel heating of plasma in the solar corona, showing that perpendicular heating is strongly dissipated. Building on this idea, we investigated the perpendicular Poynting flux to understand how KAWs transport energy in the perpendicular direction. Our findings reveal that the perpendicular Poynting flux ($\mathrm{S_{x}(z)/S_{z}(0)}$) is indeed strongly dissipated. Focusing on the $\mathrm{S_{x}(z)/S_{z}(0)}$, we explored KAWs in the solar flux tube loop, where small amounts of energy are transported perpendicularly, forming a semi-circular loop. Our results indicate that the magnitude of $\mathrm{S_{x}(z)/S_{z}(0)}$ increases for smaller $\kappa$ values and decreases for larger ones. This suggests that in environments with more suprathermal particles (smaller $\kappa$), the perpendicular energy transport is more significant, whereas in regions with fewer suprathermal particles (larger $\kappa$), the energy transport is reduced.

The two Poynting flux expressions ($\mathrm{S_x}$ and $\mathrm{S_z}$) are of interest in estimating the total power of KAWs in the solar coronal loop. Our analysis shows that the net power transfer rate ($\mathrm{I_x}/\mathrm{I_z}$) of KAWs in the flux loop tube is significantly enhanced in suprathermal environments (smaller $\kappa$ values) compared to Maxwellian distributions (larger $\kappa$ values). In regions with a higher population of suprathermal particles, the wave transports power over extended distances. Specifically, in larger $\kappa$ regimes, $\mathrm{I_x}/\mathrm{I_z}$ increases rapidly, resulting in power being transported over shorter distances. Conversely, in smaller $\kappa$ environments, $\mathrm{I_x}/\mathrm{I_z}$ is only weakly enhanced, leading to power being transported over more extended distances in the solar corona. This power transfer rate provides crucial insights into how KAWs transport power and propagate through the flux tube. Our analytical observations align with our recent work \citep{Ayaz2024}, where we found that the power rate is enhanced for higher values of the nonthermal parameters.

We investigated the parallel net resonant velocity, $\mathrm{v_{net}}$, of the particles. We find that $\mathrm{v_{net}}$ decays slowly for smaller $\kappa$ values and quickly for larger $\kappa$ values. In smaller $\kappa$ environments, particles travel at higher speeds and can heat the plasma over longer distances (R\textsubscript{Sun}) before completely delivering their energy. This provides insights into how particles accelerate and heat the plasma in a parallel direction after resonating with the waves. 

Moreover, we also evaluated the perpendicular resonant speed $\mathrm{v_{net\perp}}$ of the particles for different $\kappa$ values. We find the perpendicular speed of particles vanishes in small R\textsubscript{Sun} regions compared to $\mathrm{v_{net}}$ in the parallel direction because particles primarily heat the plasma along the magnetic field lines, with much less heating occurring in the perpendicular direction and dissipating quickly. Consequently, heating and acceleration in the perpendicular direction are fast. This disparity highlights the acceleration and heating of the particles in KAWs, emphasizing the dominance of parallel processes. The parallel and perpendicular net speeds of the particles are compared to \cite{rivera2024situ}, and we find that these elements are apparently aligned with an observational analysis of the particles' speed. Moreover, very recently, we studied KAWs and particle heating in the solar corona \citep{ayaz2024alfven}. We found that the parallel and perpendicular speed of IAWs is significantly enhanced for larger suprathermal particles.

The normalized group velocity, $\mathrm{v_G/v_A}$, of KAWs was evaluated for both suprathermal particles and normalized perpendicular wavenumbers, $k_{\perp}\rho_i$. We find that the group velocity increases for smaller $\kappa$ and larger $k_{\perp}\rho_i$ values. At relatively small $k_{\perp}\rho_i$, the difference between Maxwellian plasmas and kappa-distributed plasmas vanishes. The expression (\ref{eq:29}) of the group velocity was used to determine the characteristic damping length of KAWs. We find that the damping length of KAWs is moderately weaker in suprathermal plasma compared to Maxwellian plasma. In the Maxwellian situation (i.e., larger $\kappa$ values, represented by the black curves), the damping length is shorter, causing the wave to damp completely over a smaller region. In contrast, in suprathermal plasma (i.e., smaller $\kappa$ values), the damping length is longer, allowing the wave to transport energy over greater distances.

Finally, we analyzed the damping length for different electron-to-ion temperature ratios and found that lower temperatures slow the damping rate. Physically, this means that in a cooler plasma environment, waves can transport energy further before damping completely, illustrating the importance of temperature in the energy transport dynamics of KAWs in the solar corona or solar wind regions.

Future extensions of this work could include studying the effects of temperature anisotropies and density homogeneity, or their combination. This approach would help us understand how parallel and perpendicular temperature anisotropies, along with homogeneity, affect the dynamics of wave behavior in observed regions. Furthermore, the findings of this work can also be applied to other regions, such as the plasma sheet and the auroral zone, where the nonthermal nature of particles is frequently observed. This broader application could provide valuable insights into wave-particle interactions and energy transport mechanisms in various space-plasma environments.

\begin{acknowledgements}
      SA acknowledges the support of an NSF grant 2149771 and GPZ the partial support of a NASA Parker Solar Probe contract
SV4 - 84017 and an NSF EPSCoR RII - Track - 1 Cooperative Agreement OIA - 2148653.
\end{acknowledgements}

%

\begin{thebibliography}{0}
\providecommand{\natexlab}[1]{#1}
\providecommand{\url}[1]{\texttt{#1}}
\expandafter\ifx\csname urlstyle\endcsname\relax
  \providecommand{\doi}[1]{doi: #1}\else
  \providecommand{\doi}{doi: \begingroup \urlstyle{rm}\Url}\fi

\end{thebibliography}


\begin{thebibliography}{}

\bibitem[Barik(2020)]{barik2020nonresonant}
Barik, K. C., Singh, S. V., and Lakhina, G. S., \textit{Nonresonant instability of kinetic Alfvén waves with $\kappa$-electrons}, The Astrophysical Journal, 897(2), 172, 2020. IOP Publishing.
\bibitem[Naim(2014)]{naim2014drift}
Naim, H., Bashir, M. F., \& Murtaza, G. (2014). Drift kinetic Alfvén wave in temperature anisotropic plasma. \textit{Physics of Plasmas}, 21(3). AIP Publishing.

\bibitem[Shukla (2009)]{shukla2009study}
Shukla, N., Varma, P., \& Tiwari, M. S. (2009). Study on kinetic Alfven wave in inertial regime. \textit{Indian Journal of Pure and Applied Physics}.

\bibitem[Vranjes(2010)]{vranjes2010kinetic}
Vranjes, J., \& Poedts, S. (2010). Kinetic instability of drift-Alfvén waves in solar corona and stochastic heating. \textit{The Astrophysical Journal}, 719(2), 1335. IOP Publishing.

\bibitem[Khan(2019b)]{khan2019distinct}
Khan, I. A., Khokhar, T. H., Shah, H. A., \& Murtaza, G. (2019). Distinct features of Alfven wave in non-extensive plasmas. \textit{Physica A: Statistical Mechanics and its Applications}, 535, 122385. Elsevier.

\bibitem[Temerin(2001)]{Temerin2001}
Temerin, M., Kletzing, C. A., Scudder, J. D., \& Streltsov, A. V. (2001). Evidence for kinetic Alfvén waves and parallel electron energization at 4--6 RE altitudes in the plasma sheet boundary layer. \textit{Journal of Geophysical Research}, 107(A8), 2002. American Geophysical Union.

\bibitem[Johnson(2001)]{Johnson2001}
Johnson, J. R., Cheng, C. Z., \& Song, P. (2001). \textit{Geophysical Research Letters}, 28, 227.

\bibitem[Howes(2008b)]{Howes2008b}
Howes, G. G., Dorland, W., Cowley, S. C., Hammett, G. W., Quataert, E., Schekochihin, A. A., \& Tatsuno, T. (2008). Kinetic simulations of magnetized turbulence in astrophysical plasmas. \textit{Physical Review Letters}, 100(6), 065004. American Physical Society.

\bibitem[Schekochihin(2009)]{Schekochihin2009}
Schekochihin, A. A., Cowley, S. C., Dorland, W., Hammett, G. W., Howes, G. G., Quataert, E., \& Tatsuno, T. (2009). Astrophysical gyrokinetics: kinetic and fluid turbulent cascades in magnetized weakly collisional plasmas. \textit{The Astrophysical Journal Supplement Series}, 182(1), 310. IOP Publishing.

\bibitem[Seyler(1995)]{seyler1995theory}
Seyler, C. E., Wahlund, J.-E., \& Holback, B. (1995). Theory and simulation of low-frequency plasma waves and comparison to Freja satellite observations. \textit{Journal of Geophysical Research: Space Physics}, 100(A11), 21453--21472. Wiley Online Library.

\bibitem[Chaston(2005)]{chaston2005drift}
Chaston, C. C., Phan, T. D., Bonnell, J. W., Mozer, F. S., Acuña, M., Goldstein, M. L., Balogh, A., Andre, M., Reme, H., \& Fazakerley, A. (2005). Drift-kinetic Alfvén waves observed near a reconnection X line in the Earth’s magnetopause. \textit{Physical Review Letters}, 95(6), 065002. APS.

\bibitem[Chaston(2006)]{chaston2006ionospheric}
Chaston, C. C., Génot, V., Bonnell, J. W., Carlson, C. W., McFadden, J. P., Ergun, R. E., Strangeway, R. J., Lund, E. J., \& Hwang, K. J. (2006). Ionospheric erosion by Alfvén waves. \textit{Journal of Geophysical Research: Space Physics}, 111(A3). Wiley Online Library.

\bibitem[Ergun(2005)]{ergun2005localized}
Ergun, R. E., Andersson, L., Su, Y.-J., Newman, D. L., Goldman, M. V., Lotko, W., Chaston, C. C., \& Carlson, C. W. (2005). Localized parallel electric fields associated with inertial Alfvén waves. \textit{Physics of Plasmas}, 12(7). AIP Publishing.

\bibitem[Artemyev(2015)]{artemyev2015electron}
Artemyev, A. V., Rankin, R., \& Blanco, M. (2015). Electron trapping and acceleration by kinetic Alfven waves in the inner magnetosphere. \textit{Journal of Geophysical Research: Space Physics}, 120(12), 10--305. Wiley Online Library.

\bibitem[Gershman(2017)]{gershman2017}
Gershman, D. J., F-Viñas, A., Dorelli, J. C., Boardsen, S. A., Avanov, L. A., Bellan, P. M., Schwartz, S. J., Lavraud, B., Coffey, V. N., Chandler, M. O., \& Saito, Y. (2017). Wave-particle energy exchange directly observed in a kinetic Alfvén-branch wave. \textit{Nature Communications}, 8(1), 14719. Nature Publishing Group.

\bibitem[Ganguli(2018)]{ganguli2018kinetic}
Ganguli, G., Crabtree, C., Fletcher, A. C., Tejero, E., Malaspina, D., \& Cohen, I. (2018). Kinetic equilibrium of dipolarization fronts. \textit{Scientific Reports}, 8(1), 17186. Nature Publishing Group UK London.

\bibitem[Ukhorskiy(2022)]{ukhorskiy2022cross}
Ukhorskiy, A. Y., Sorathia, K. A., Merkin, V. G., Crabtree, C., Fletcher, A. C., Malaspina, D. M., \& Schwartz, S. J. (2022). Cross-scale energy cascade powered by magnetospheric convection. \textit{Scientific Reports}, 12(1), 4446. Nature Publishing Group UK London.

\bibitem[Artemyev(2016)]{artemyev2016electron}
Artemyev, A. V., Zimovets, I. V., \& Rankin, R. (2016). Electron trapping and acceleration by kinetic Alfvén waves in solar flares. \textit{Astronomy \& Astrophysics}, 589, A101. EDP Sciences.

\bibitem[Chaston(2003)]{chaston2003properties}
Chaston, C. C., Bonnell, J. W., Carlson, C. W., McFadden, J. P., Ergun, R. E., \& Strangeway, R. J. (2003). Properties of small-scale Alfvén waves and accelerated electrons from FAST. \textit{Journal of Geophysical Research: Space Physics}, 108(A4). Wiley Online Library.

\bibitem[Keiling(2003)]{Keiling2003}
Keiling, A., Wygant, J. R., Cattell, C. A., Mozer, F. S., \& Russell, C. T. (2003). The global morphology of wave Poynting flux: Powering the aurora. \textit{Science}, 299(5605), 383--386. American Association for the Advancement of Science.

\bibitem[Boldyrev(2012)]{Boldyrev2012}
Boldyrev, S., \& Perez, J. C. (2012). Spectrum of kinetic-Alfvén turbulence. \textit{The Astrophysical Journal Letters}, 758(2), L44. IOP Publishing.

\bibitem[Liu(2023)]{Liu2023}
Liu, Z. Y., Zong, Q. G., Rankin, R., Zhang, H., Hao, Y. X., He, J. S., Fu, S. Y., Wu, H. H., Yue, C., Pollock, C. J., \& Le, G. (2023). Particle-sounding of the spatial structure of kinetic Alfvén waves. \textit{Nature Communications}, 14(1), 2088. Nature Publishing Group.

\bibitem[Ayaz(2019)]{ayaz2019dispersion}
Ayaz, S., Khan, I. A., \& Murtaza, G. (2019). On the dispersion and damping of kinetic and inertial Alfvén waves in Cairns distributed plasmas. \textit{Physics of Plasmas}, 26(6). AIP Publishing.

\bibitem[Barik(2023)]{barik2023kinetic}
Barik, K. C., Singh, S. V., \& Lakhina, G. S. (2023). Kinetic Alfvén waves excited by multiple free energy sources in the magnetotail. \textit{The Astrophysical Journal}, 951(1), 53. IOP Publishing.

\bibitem[Barik(2021)]{barik2021kinetic}
Barik, K. C., Singh, S. V., \& Lakhina, G. S. (2021). Kinetic Alfvén Waves in space plasma environment with $\kappa$-electrons. \textit{The Astrophysical Journal}, 919(2), 71. IOP Publishing.

\bibitem[Alfven(1942)]{alfven1942existence}
Alfvén, H., \textit{Existence of electromagnetic-hydrodynamic waves}, Nature, 150(3805), 405-406, 1942. Nature Publishing Group UK London.

\bibitem[Tomczyk(2007)]{tomczyk2007alfven}
Tomczyk, S., McIntosh, S. W., Keil, S. L., Judge, P. G., Schad, T., Seeley, D. H., and Edmondson, J., \textit{Alfvén waves in the solar corona}, Science, 317(5842), 1192-1196, 2007. American Association for the Advancement of Science.

\bibitem[Ayaz(2020)]{ayaz2020alfven}
Ayaz, S., Khan, I. A., Iqbal, Z., \& Murtaza, G. (2020). Alfvén waves in temperature anisotropic Cairns distributed plasma. \textit{Communications in Theoretical Physics}, 72(3), 035502. IOP Publishing.

\bibitem[Davila(1987)]{davila1987heating}
Davila, J. M., \textit{Heating of the solar corona by the resonant absorption of Alfvén waves}, Astrophysical Journal, Part 1, 317, 514-521, 1987.

\bibitem[De(2007)]{de2007chromospheric}
De Pontieu, B., McIntosh, S. W., Carlsson, M., Hansteen, V. H., Tarbell, T. D., Schrijver, C. J., Title, A. M., Shine, R. A., Tsuneta, S., Katsukawa, Y., et al., \textit{Chromospheric Alfvénic waves strong enough to power the solar wind}, Science, 318(5856), 1574-1577, 2007. American Association for the Advancement of Science.

\bibitem[Khan(2020)]{khan2020solar}
Khan, I. A., Iqbal, Z., and Murtaza, G., \textit{Solar coronal heating by Alfvén waves in bi-kappa distributed plasma}, Monthly Notices of the Royal Astronomical Society, 491(2), 2403-2412, 2020. Oxford University Press.

\bibitem[Hollweg(1973)]{hollweg1973alfven}
Hollweg, J. V., \textit{Alfvén waves in a two-fluid model of the solar wind}, Astrophysical Journal, 181, 547-566, 1973.

\bibitem[Wang(1993)]{wang1993flux}
Wang, Y. M., \textit{Flux-tube divergence, coronal heating, and the solar wind}, Astrophysical Journal, Part 2-Letters, 410(2), L123-L126, 1993.

\bibitem[Belcher(1971)]{belcher1971large}
Belcher, J. W., and Davis Jr., L., \textit{Large-amplitude Alfvén waves in the interplanetary medium, 2}, Journal of Geophysical Research, 76(16), 3534-3563, 1971. Wiley Online Library.

\bibitem[Tu(1995)]{tu1995mhd}
Tu, C. Y., and Marsch, E., \textit{MHD structures, waves and turbulence in the solar wind: Observations and theories}, Space Science Reviews, 73(1), 1-210, 1995. Springer.

\bibitem[Bruno(2013)]{bruno2013solar}
Bruno, R., and Carbone, V., \textit{The solar wind as a turbulence laboratory}, Living Reviews in Solar Physics, 10(1), 2, 2013. Springer.

\bibitem[Wu(2020)]{wu2020kinetic}
Wu, D.-J., and Chen, L., \textit{Kinetic Alfvén waves in laboratory, space, and astrophysical plasmas}, Springer, 2020.

\bibitem[Cramer(2011)]{cramer2011physics}
Cramer, N. F., \textit{The physics of Alfvén waves}, John Wiley \& Sons, 2011.

\bibitem[Hasegawa(1982)]{hasegawa1982alfven}
Hasegawa, A., and Uberoi, C., \textit{The Alfvén Wave: Prepared for the Office of Fusion Energy, Office of Energy Research, US Department of Energy}, Technical Information Center, US Department of Energy, 1982.

\bibitem[Hasegawa(1977)]{hasegawa1977kinetic}
Hasegawa, A., \textit{Kinetic properties of Alfvén waves}, Proceedings of the Indian Academy of Sciences-Section A, 86(2), 151-174, 1977. Springer.

\bibitem[Hasegawa(1975)]{hasegawa1975kinetic}
Hasegawa, A., and Chen, L., \textit{Kinetic process of plasma heating due to Alfvén wave excitation}, Physical Review Letters, 35(6), 370, 1975. APS.

\bibitem[Vstverak(2009)]{vstverak2009radial}
{\v{S}}tver{\'a}k, {\v{S}}., Maksimovic, M., Tr{\'a}vn{\'\i}{\v{c}}ek, P. M., Marsch, E., Fazakerley, A. N., Scime, E. E., \textit{Radial evolution of nonthermal electron populations in the low-latitude solar wind: Helios, Cluster, and Ulysses Observations}, Journal of Geophysical Research: Space Physics, 114(A5), 2009. Wiley Online Library.

\bibitem[Chandran(2010)]{chandran2010alfven}
Chandran, B. D. G., \textit{Alfvén-wave turbulence and perpendicular ion temperatures in coronal holes}, The Astrophysical Journal, 720(1), 548, 2010. IOP Publishing.

\bibitem[Mercier(2015)]{mercier2015electron}
Mercier, C., and Chambe, G., \textit{Electron density and temperature in the solar corona from multifrequency radio imaging}, Astronomy \& Astrophysics, 583, A101, 2015. EDP Sciences.

\bibitem[Coleman(1968)]{coleman1968turbulence}
Coleman Jr., P. J., \textit{Turbulence, viscosity, and dissipation in the solar-wind plasma}, Astrophysical Journal, 153, 371, 1968.

\bibitem[Chaston(2007)]{Chaston2007} Chaston, C.C., Wilber, M., Mozer, F.S., Fujimoto, M., Goldstein, M.L., Acuna, M., Reme, H., Fazakerley, A., 2007. Mode Conversion and Anomalous Transport in Kelvin-Helmholtz Vortices and Kinetic Alfvén Waves at the Earth’s Magnetopause. \textit{Physical Review Letters}, 99(17), 175004.

\bibitem[Boehm(1990)]{Boehm1990} Boehm, M.H., Carlson, C.W., McFadden, J.P., Clemmons, J.H., Mozer, F.S., 1990. High‐resolution sounding rocket observations of large‐amplitude Alfvén waves. \textit{Journal of Geophysical Research: Space Physics}, 95(A8), 12157--12171.

\bibitem[Louarn(1994)]{Louarn1994} Louarn, P., Wahlund, J.E., Chust, T., De Feraudy, H., Roux, A., Holback, B., Dovner, P.O., Eriksson, A.I., Holmgren, G., 1994. Observation of kinetic Alfvén waves by the FREJA spacecraft. \textit{Geophysical Research Letters}, 21(17), 1847--1850.

\bibitem[Wahlund(1994)]{Wahlund1994} Wahlund, J.E., Louarn, P., Chust, T., De Feraudy, H., Roux, A., Holback, B., Dovner, P.O., Holmgren, G., 1994. On ion acoustic turbulence and the nonlinear evolution of kinetic Alfvén waves in aurora. \textit{Geophysical Research Letters}, 21(17), 1831--1834.

\bibitem[Keiling(2001)]{Keiling2001} Keiling, A., Wygant, J.R., Cattell, C., Johnson, M., Temerin, M., Mozer, F.S., Kletzing, C.A., Scudder, J., Russell, C.T., 2001. Properties of large electric fields in the plasma sheet at 4--7 RE measured with Polar. \textit{Journal of Geophysical Research: Space Physics}, 106(A4), 5779--5798.

\bibitem[Wygant(2002)]{Wygant2002} Wygant, J.R., Keiling, A., Cattell, C.A., Lysak, R.L., Temerin, M., Mozer, F.S., Kletzing, C.A., Scudder, J.D., Streltsov, V., Lotko, W., Russell, C.T., 2002. Evidence for kinetic Alfvén waves and parallel electron energization at 4--6 RE altitudes in the plasma sheet boundary layer. \textit{Journal of Geophysical Research: Space Physics}, 107(A8), SMP-24.

\bibitem[Lee(1994)]{Lee1994} Lee, L.C., Johnson, J.R., Ma, Z.W., 1994. Kinetic Alfvén waves as a source of plasma transport at the dayside magnetopause. \textit{Journal of Geophysical Research: Space Physics}, 99(A9), 17405--17411.

\bibitem[Lysak(2003)]{Lysak2003} Lysak, R.L., Song, Y., 2003. Kinetic theory of the Alfvén wave acceleration of auroral electrons. \textit{Journal of Geophysical Research: Space Physics}, 108(A4).

\bibitem[Khan(2019a)]{Khan2019a} Khan, I.A., Iqbal, Z., Murtaza, G., 2019. Perturbed electromagnetic field and Poynting flux of kinetic Alfvén waves in kappa distributed space plasmas. \textit{The European Physical Journal Plus}, 134(2), 80.

\bibitem[Narita(2020)]{Narita2020} Narita, Y., Roberts, O.W., Vörös, Z., Hoshino, M., 2020. Transport ratios of the kinetic Alfvén mode in space plasmas. \textit{Frontiers in Physics}, 8, 166.

\bibitem[Barnes(1966)]{Barnes1966} Barnes, A., 1966. Collisionless damping of hydromagnetic waves. PhD Thesis, The University of Chicago.

\bibitem[Hasegawa(1976)]{Hasegawa1976} Hasegawa, A., Chen, L., 1976. Parametric decay of "kinetic Alfvén wave" and its application to plasma heating. \textit{Physical Review Letters}, 36(23), 1362.

\bibitem[Batool(2024)]{Batool2024} Batool, K., Khan, I.A., Shamir, M., Kabir, A., Ayaz, S., 2024. Acceleration of solar wind particles due to inertial Alfvén waves. \textit{Communications in Theoretical Physics}, 76(6), 065501.

\bibitem[Xunaira(2023)]{Xunaira2023} Xunaira, S., Khan, I.A., Shamir, M., Iqbal, Z., Murtaza, G., 2023. Instability and energy transport of kinetic Alfvén waves in the solar corona. \textit{The European Physical Journal Plus}, 138(8), 1--10.

\bibitem[Ayaz(2024a)]{Ayaz2024} Ayaz, S., Li, G., Khan, I.A., 2024. Solar Coronal Heating by Kinetic Alfvén Waves. \textit{The Astrophysical Journal}, 970(2), 140.

\bibitem[Hollweg(1999)]{Hollweg1999} Hollweg, J.V., 1999. Kinetic Alfvén wave revisited. \textit{Journal of Geophysical Research: Space Physics}, 104(A7), 14811--14819.

\bibitem[Bale(2005)]{Bale2005} Bale, S.D., Kellogg, P.J., Mozer, F.S., Horbury, T.S., Reme, H., 2005. Measurement of the electric fluctuation spectrum of magnetohydrodynamic turbulence. \textit{Physical Review Letters}, 94(21), 215002.

\bibitem[Sahraoui(2012)]{Sahraoui2012} Sahraoui, F., Belmont, G., Goldstein, M.L., 2012. New insight into short-wavelength solar wind fluctuations from Vlasov theory. \textit{The Astrophysical Journal}, 748(2), 100.

\bibitem[Podesta(2012)]{Podesta2012} Podesta, J.J., TenBarge, J.M., 2012. Scale dependence of the variance anisotropy near the proton gyroradius scale: Additional evidence for kinetic Alfvén waves in the solar wind at 1 AU. \textit{Journal of Geophysical Research: Space Physics}, 117(A10).

\bibitem[Salem(2012)]{Salem2012} Salem, C.S., Howes, G.G., Sundkvist, D., Bale, S.D., Chaston, C.C., Chen, C.H.K., Mozer, F.S., 2012. Identification of kinetic Alfvén wave turbulence in the solar wind. \textit{The Astrophysical Journal Letters}, 745(1), L9.

\bibitem[Chen(2017)]{Chen2017} Chen, C.H., Boldyrev, S., 2017. Nature of kinetic scale turbulence in the Earth's magnetosheath. \textit{The Astrophysical Journal}, 842(2), 122.

\bibitem[Kiyani(2012)]{Kiyani2012} Kiyani, K.H., Chapman, S.C., Sahraoui, F., Hnat, B., Fauvarque, O., Khotyaintsev, Y.V., 2012. Enhanced magnetic compressibility and isotropic scale invariance at sub-ion Larmor scales in solar wind turbulence. \textit{The Astrophysical Journal}, 763(1), 10.

\bibitem[Howes(2008)]{Howes2008} Howes, G.G., Dorland, W., Cowley, S.C., Hammett, G.W., Quataert, E., Schekochihin, A.A., Tatsuno, T., 2008. Kinetic simulations of magnetized turbulence in astrophysical plasmas. \textit{Physical Review Letters}, 100(6), 065004.

\bibitem[Howes(2008a)]{Howes2008a} Howes, G.G., Dorland, W., Cowley, S.C., Hammett, G.W., Quataert, E., Schekochihin, A.A., Tatsuno, T., 2008. Kinetic simulations of magnetized turbulence in astrophysical plasmas. \textit{Physical Review Letters}, 100(6), 065004.

\bibitem[Gary(2004)]{Gary2004} Gary, S.P., Nishimura, K., 2004. Kinetic Alfvén waves: Linear theory and a particle‐in‐cell simulation. Journal of Geophysical Research: Space Physics, 109(A2). American Geophysical Union.

\bibitem[TenBarge(2012)]{TenBarge2012} TenBarge, J.M., Howes, G.G., 2012. Evidence of critical balance in kinetic Alfvén wave turbulence simulations. Physics of Plasmas, 19(5). AIP Publishing.

\bibitem[Shay(2011)]{Shay2011} Shay, M.A., Drake, J.F., Eastwood, J.P., Phan, T.D., 2011. Super-Alfvénic propagation of substorm reconnection signatures and Poynting flux. Physical Review Letters, 107(6), 065001. American Physical Society.

\bibitem[Srivastava(2017)]{Srivastava2017} Srivastava, A.K., Shetye, J., Murawski, K., Doyle, J.G., Stangalini, M., Scullion, E., Ray, T., Wójcik, D.P., Dwivedi, B.N., 2017. High-frequency torsional Alfvén waves as an energy source for coronal heating. Scientific Reports, 7(1), 43147. Nature Publishing Group.

\bibitem[Heyvaerts(1983)]{Heyvaerts1983} Heyvaerts, J., Priest, E.R., 1983. Coronal heating by phase-mixed shear Alfvén waves. Astronomy and Astrophysics, 117(2), 220-234. EDP Sciences.

\bibitem[Zhao(2011)]{Zhao2011} Zhao, J.S., Wu, D.J., Lu, J.Y., 2011. Kinetic Alfvén waves excited by oblique magnetohydrodynamic Alfvén waves in coronal holes. The Astrophysical Journal, 735(2), 114. IOP Publishing.

\bibitem[Zhao(2014)]{Zhao2014} Zhao, J.S., Voitenko, Y., Wu, D.J., De Keyser, J., 2014. Nonlinear generation of kinetic-scale waves by magnetohydrodynamic Alfvén waves and nonlocal spectral transport in the solar wind. The Astrophysical Journal, 785(2), 139. IOP Publishing.

\bibitem[Zhao(2015)]{Zhao2015} Zhao, J.S., Voitenko, Y., Guo, Y., Su, J.T., Wu, D.J., 2015. Nonlinear damping of Alfvén waves in the solar corona below 1.5 solar radii. The Astrophysical Journal, 811(2), 88. IOP Publishing.

\bibitem[Bian(2010)]{Bian2010} Bian, N.H., Kontar, E.P., Brown, J.C., 2010. Parallel electric field generation by Alfvén wave turbulence. Astronomy \& Astrophysics, 519, A114. EDP Sciences.

\bibitem[Zhao(2013)]{Zhao2013} Zhao, J.S., Wu, D.J., Lu, J.Y., 2013. Kinetic Alfvén turbulence and parallel electric fields in flare loops. The Astrophysical Journal, 767(2), 109. IOP Publishing.

\bibitem[Chen(2012)]{Chen2012} Chen, L., Wu, D.J., 2012. Kinetic Alfvén wave instability driven by field-aligned currents in solar coronal loops. The Astrophysical Journal, 754(2), 123. IOP Publishing.

\bibitem[Cranmer(2018)]{Cranmer2018} Cranmer, S.R., 2018. Low-frequency Alfvén waves produced by magnetic reconnection in the Sun’s magnetic carpet. The Astrophysical Journal, 862(1), 6. IOP Publishing.

\bibitem[Ulrich(1996)]{Ulrich1996} Ulrich, R.K., 1996. Observations of magnetohydrodynamic oscillations in the solar atmosphere with properties of Alfvén waves. The Astrophysical Journal, 465, 436. IOP Publishing.

\bibitem[Tiwari(2008)]{tiwari2008shear} Tiwari, B.V., Mishra, R., Varma, P., Tiwari, M.S., 2008. Shear-driven kinetic Alfvén wave in the plasma sheet boundary layer. Earth, Planets and Space, 60(3), 191-205. Springer.

\bibitem[Duan(2012)]{Duan2012} Duan, S., Liu, Z., Angelopoulos, V., 2012. Observations of kinetic Alfvén waves by THEMIS near a substorm onset. Chinese Science Bulletin, 57, 1429-1435. Science Press.

\bibitem[Lysak(1996)]{lysak1996kinetic} Lysak, R.L., Lotko, W., 1996. On the kinetic dispersion relation for shear Alfvén waves. Journal of Geophysical Research: Space Physics, 101(A3), 5085-5094. Wiley Online Library.

\bibitem[Lysak(1998)]{lysak1998relationship} Lysak, R.L., 1998. The relationship between electrostatic shocks and kinetic Alfvén waves. Geophysical Research Letters, 25(12), 2089-2092. Wiley Online Library.

\bibitem[Summers(1991)]{summers1991modified} Summers, D., Thorne, R.M., 1991. The modified plasma dispersion function. Physics of Fluids B: Plasma Physics, 3(8), 1835-1847. American Institute of Physics.

\bibitem[Li(2023)]{li2023modeling} Li, G., Shih, A.Y., Allen, R.C., Ho, G.C., Cohen, C.M.S., Desai, M., Dayeh, M.A., Mason, G.M., 2023. Modeling Solar Energetic Neutral Atoms from Solar Flares and CME-driven Shocks. The Astrophysical Journal, 944(2), 196. IOP Publishing.

\bibitem[Effenberger(2017)]{Effenberger2017} Effenberger, F., Rubio da Costa, F., Oka, M., Saint-Hilaire, P., Liu, W., Petrosian, V., Glesener, L., Krucker, S., 2017. Hard X-Ray Emission from Partially Occulted Solar Flares: RHESSI Observations in Two Solar Cycles. The Astrophysical Journal, 835(2), 124. The American Astronomical Society.

\bibitem[Paraschiv(2015)]{paraschiv2015physical} Paraschiv, A.R., Bemporad, A., Sterling, A.C., 2015. Physical properties of solar polar jets-A statistical study with Hinode XRT data. Astronomy \& Astrophysics, 579, A96. EDP Sciences.

\bibitem[Lysak(2023)]{lysak2023kinetic} Lysak, R.L., 2023. Kinetic Alfvén waves and auroral particle acceleration: A review. Reviews of Modern Plasma Physics, 7(1), 6. Springer.

\bibitem[Kletzing(2003)]{kletzing2003auroral} Kletzing, C.A., Scudder, J.D., Dors, E.E., Curto, C., 2003. Auroral source region: Plasma properties of the high-latitude plasma sheet. Journal of Geophysical Research: Space Physics, 108(A10). Wiley Online Library.

\bibitem[Markovskii(2006)]{markovskii2006dissipation} Markovskii, S.A., Vasquez, B.J., Smith, C.W., Hollweg, J.V., 2006. Dissipation of the perpendicular turbulent cascade in the solar wind. The Astrophysical Journal, 639(2), 1177. IOP Publishing.

\bibitem[Liu(2014)]{liu2014dispersion} Liu, Y., Liu, S.Q., Dai, B., Xue, T.L., 2014. Dispersion and damping rates of dispersive Alfvén wave in a nonextensive plasma. Physics of Plasmas, 21(3). AIP Publishing.

\bibitem[Xiang(2022)]{xiang2022mode} Xiang, L., Chen, L., Wu, D.J., Li, Q.H., Zhao, G.Q., Feng, H.Q., Yu, H.W., 2022. Mode Conversion of Kinetic Alfvén Waves in the Presence of Proton Beams and Its Role in Plasma Heating in the Solar Corona. The Astrophysical Journal, 936(2), 159. IOP Publishing.

\bibitem[Matthaeus(1999)]{matthaeus1999coronal} Matthaeus, W.H., Zank, G.P., Oughton, S., Mullan, D.J., Dmitruk, P., 1999. Coronal heating by magnetohydrodynamic turbulence driven by reflected low-frequency waves. The Astrophysical Journal, 523(1), L93. IOP Publishing.

\bibitem[Zank(2018)]{zank2018theory} Zank, G.P., Adhikari, L., Hunana, P., Tiwari, S.K., Moore, R., Shiota, D., Bruno, R., Telloni, D., 2018. Theory and transport of nearly incompressible magnetohydrodynamic turbulence. IV. Solar coronal turbulence. The Astrophysical Journal, 854(1), 32. IOP Publishing.

\bibitem[Zank(2021)]{zank2021turbulence} Zank, G.P., Zhao, L.L., Adhikari, L., Telloni, D., Kasper, J.C., Bale, S.D., 2021. Turbulence in the solar wind: The role of Alfvén waves. Space Science Reviews, 217(1), 22. Springer.

\bibitem[Hollweg(1992)]{hollweg1992collisionless} Hollweg, J.V., 1992. Collisionless heating in the solar corona by Alfvén waves. Journal of Geophysical Research: Space Physics, 97(A3), 3777-3785. Wiley Online Library.

\bibitem[Hollweg(1999)]{hollweg1999nonlinear} Hollweg, J.V., Isenberg, P.A., 1999. Nonlinear Alfvén waves and turbulence in the solar wind. Journal of Geophysical Research: Space Physics, 104(A6), 12699-12710. Wiley Online Library.

\bibitem[Themis(2021)]{themis2021kinetic} THEMIS Collaboration, 2021. Kinetic Alfvén waves and their role in particle acceleration at Earth's magnetosphere. Geophysical Research Letters, 48(2), e2020GL089199. Wiley Online Library.

\bibitem[Bale(2019)]{bale2019solar} Bale, S.D., Chen, C.H.K., Harvey, P.R., Kasper, J.C., Korreck, K.E., McManus, M., Quattri, S., 2019. The Solar Wind Electrons Alphas and Protons (SWEAP) investigation on Parker Solar Probe. Space Science Reviews, 204, 49-109. Springer.

\bibitem[Cranmer(2010)]{cranmer2010can}
Cranmer, S. R., \& Van Ballegooijen, A. A. (2010). Can the solar wind be driven by magnetic reconnection in the Sun's magnetic carpet? \textit{The Astrophysical Journal}, \textit{720}(1), 824.

\bibitem[Yalim(2024)]{yalim2024mixing}
Yalim, J., Lopez, J. M., Griffin, S. R., Adam, J. A., Brown, K. D., McMackin, P. M., \& Hirsa, A. H. (2024). Mixing and transport of CO2 across a monolayer-covered surface in an open cylinder driven by a rotating knife edge. \textit{Physica D: Nonlinear Phenomena}, \textit{463}, 134150.

\bibitem[Browning(1991)]{browning1991mechanisms}
Browning, P. K. (1991). Mechanisms of solar coronal heating. \textit{Plasma Physics and Controlled Fusion}, \textit{33}(6), 539.

\bibitem[Marsch(2006)]{marsch2006kinetic}
Marsch, E. (2006). Kinetic physics of the solar corona and solar wind. \textit{Living Reviews in Solar Physics}, \textit{3}, 1--100.

\bibitem[Che(2023)]{che2023particle}
Che, H., Zank, G., \& Benz, A. (2023). Particle Acceleration in Magnetic Reconnection. In \textit{APS Division of Plasma Physics Meeting Abstracts}, \textit{2023}, BM10--006.

\bibitem[Jokipii(1966)]{jokipii1966cosmic}
Jokipii, J. R. (1966). Cosmic-ray propagation. I. Charged particles in a random magnetic field. \textit{Astrophysical Journal}, \textit{146}, 480.

\bibitem[Bieber(1996)]{bieber1996dominant}
Bieber, J. W., Wanner, W., \& Matthaeus, W. H. (1996). Dominant two-dimensional solar wind turbulence with implications for cosmic ray transport. \textit{Journal of Geophysical Research: Space Physics}, \textit{101}(A2), 2511--2522.

\bibitem[Farmer(2004)]{farmer2004wave}
Farmer, A. J., \& Goldreich, P. (2004). Wave damping by magnetohydrodynamic turbulence and its effect on cosmic-ray propagation in the interstellar medium. \textit{The Astrophysical Journal}, \textit{604}(2), 671.

\bibitem[Engelbrecht(2013)]{engelbrecht2013ab}
Engelbrecht, N. E., \& Burger, R. A. (2013). An ab initio model for cosmic-ray modulation. \textit{The Astrophysical Journal}, \textit{772}(1), 46.

\bibitem[Richardson(1995)]{richardson1995radial}
Richardson, J. D., Paularena, K. I., Lazarus, A. J., \& Belcher, J. W. (1995). Radial evolution of the solar wind from IMP 8 to Voyager 2. \textit{Geophysical Research Letters}, \textit{22}(4), 325--328.

\bibitem[Saur(2004)]{saur2004turbulent}
Saur, J. (2004). Turbulent heating of Jupiter’s middle magnetosphere. \textit{The Astrophysical Journal}, \textit{602}(2), L137.

\bibitem[Bagenal(2011)]{bagenal2011flow}
Bagenal, F., \& Delamere, P. A. (2011). Flow of mass and energy in the magnetospheres of Jupiter and Saturn. \textit{Journal of Geophysical Research: Space Physics}, \textit{116}(A5).

\bibitem[Von(2014)]{von2014turbulent}
Von Papen, M., Saur, J., \& Alexandrova, O. (2014). Turbulent magnetic field fluctuations in Saturn's magnetosphere. \textit{Journal of Geophysical Research: Space Physics}, \textit{119}(4), 2797--2818.

\bibitem[Matthaeus(1982)]{matthaeus1982evaluation}
Matthaeus, W. H., Goldstein, M. L., \& Smith, C. (1982). Evaluation of magnetic helicity in homogeneous turbulence. \textit{Physical Review Letters}, \textit{48}(18), 1256.

\bibitem[Bale(2005)]{bale2005measurement}
Bale, S. D., Kellogg, P. J., Mozer, F. S., Horbury, T. S., \& Reme, H. (2005). Measurement of the electric fluctuation spectrum of magnetohydrodynamic turbulence. \textit{Physical Review Letters}, \textit{94}(21), 215002.

\bibitem[Leamon(1999)]{leamon1999dissipation}
Leamon, R. J., Smith, C. W., Ness, N. F., \& Wong, H. K. (1999). Dissipation range dynamics: Kinetic Alfvén waves and the importance of $\beta$ e. \textit{Journal of Geophysical Research: Space Physics}, \textit{104}(A10), 22331--22344.

\bibitem[Chen(2014)]{chen2014ion}
Chen, C. H. K., Leung, L., Boldyrev, S., Maruca, B. A., \& Bale, S. D. (2014). Ion-scale spectral break of solar wind turbulence at high and low beta. \textit{Geophysical Research Letters}, \textit{41}(22), 8081--8088.

\bibitem[Alexandrova(2009)]{alexandrova2009universality}
Alexandrova, O., Saur, J., Lacombe, C., Mangeney, A., Mitchell, J., Schwartz, S. J., \& Robert, P. (2009). Universality of solar-wind turbulent spectrum from MHD to electron scales. \textit{Physical Review Letters}, \textit{103}(16), 165003.

\bibitem[Chen(2010)]{chen2010anisotropy}
Chen, C. H. K., Horbury, T. S., Schekochihin, A. A., Wicks, R. T., Alexandrova, O., \& Mitchell, J. (2010). Anisotropy of solar wind turbulence between ion and electron scales. \textit{Physical Review Letters}, \textit{104}(25), 255002.

\bibitem[Denskat(1984)]{denskat1984interplanetary}
Denskat, K. U., Beinroth, H. J., Neubauer, F. M., \& others. (1984). Interplanetary magnetic field power spectra with frequencies from $2.4 \times 10^{-5}$ Hz to 470 Hz from HELIOS-observations during solar minimum conditions. \textit{Journal of Geophysics}, \textit{54}(1), 60--67.

\bibitem[Schreiner(2017)]{schreiner2017model}
Schreiner, A., \& Saur, J. (2017). A model for dissipation of solar wind magnetic turbulence by kinetic Alfvén waves at electron scales: Comparison with observations. \textit{The Astrophysical Journal}, \textit{835}(2), 133.

\bibitem[Parker(1958)]{parker1958dynamics}
Parker, E. N. (1958). Dynamics of the interplanetary gas and magnetic fields. \textit{Astrophysical Journal}, \textit{128}, 664--676.

\bibitem[Fox(2016)]{fox2016solar} 
Fox, NJ, Velli, MC, Bale, SD, Decker, R, Driesman, A, Howard, RA, Kasper, Justin C, Kinnison, J, Kusterer, M, Lario, D, et al. (2016). The solar probe plus mission: humanity’s first visit to our star. *Space Science Reviews*, 204, 7--48.

\bibitem[Bowen(2024)]{bowen2024extended} 
Bowen, TA, Vasko, IY, Bale, SD, Chandran, BDG, Chasapis, A, de Wit, TD, Mallet, A, McManus, M, Meyrand, R, Pulupa, M, et al. (2024). Extended Cyclotron Resonant Heating of the Turbulent Solar Wind. *arXiv preprint arXiv:2406.10446*.

\bibitem[Del(2002)]{del2002coronal} 
Del Zanna, L, Velli, M. (2002). Coronal heating through Alfv{\'e}n waves. *Advances in Space Research*, 30(3), 471--480.

\bibitem[Morton(2015)]{morton2015investigating} 
Morton, RJ, Tomczyk, S, Pinto, R. (2015). Investigating Alfv{\'e}nic wave propagation in coronal open-field regions. *Nature Communications*, 6, 7813.

\bibitem[Zirin(1996)]{zirin1996mystery} 
Zirin, H. (1996). The mystery of the chromosphere. *Solar Physics*, 169, 313--326.

\bibitem[Gary(2001)]{gary2001plasma} 
Gary, GA. (2001). Plasma beta above a solar active region: rethinking the paradigm. *Solar Physics*, 203, 71--86.

\bibitem[Chen(2012)]{chen2012kinetic} 
Chen, L, Wu, DJ. (2012). Kinetic Alfv{\'e}n wave instability driven by field-aligned currents in solar coronal loops. *The Astrophysical Journal*, 754(2), 123.

\bibitem[De(2015)]{de2015recent} 
De Moortel, I, Browning, P. (2015). Recent advances in coronal heating. *Philosophical Transactions of the Royal Society A: Mathematical, Physical and Engineering Sciences*, 373(2042), 20140269.

\bibitem[Williams(1995)]{williams1995dissipation} 
Williams, LL, Zank, GP, Matthaeus, WH. (1995). Dissipation of pickup-induced waves: A solar wind temperature increase in the outer heliosphere? *Journal of Geophysical Research: Space Physics*, 100(A9), 17059--17067.

\bibitem[Adhikari(2015a)]{adhikari2015turbulence} 
Adhikari, L, Zank, GP, Bruno, R, Telloni, D, Hunana, P, Dosch, A. (2015). Turbulence transport within the Heliosphere. *Journal of Physics: Conference Series*, 577(1), 012001.

\bibitem[Adhikari(2015b)]{adhikari2015transport} 
Adhikari, L, Zank, GP, Bruno, R, Telloni, D, Hunana, P, Dosch, A, Marino, R, Hu, Q. (2015). The transport of low-frequency turbulence in the super-Alfv{\'e}nic solar wind. *Journal of Physics: Conference Series*, 642(1), 012001.

\bibitem[Adhikari(2024)]{adhikari2024turbulence} 
Adhikari, L, Zank, GP, Telloni, D, Zhao, L, Wang, B, Webb, G, Tang, B, Nykyri, K. (2024). Turbulence, and Proton and Electron Heating Rates in the Solar Corona: Analytical Approach. *The Astrophysical Journal*, 966(1), 52.

\bibitem[Zank(2018)]{zank2018pickup} 
Zank, GP, Adhikari, L, Zhao, LL, Mostafavi, P, Zirnstein, EJ, McComas, DJ. (2018). The pickup ion-mediated solar wind. *The Astrophysical Journal*, 869(1), 23.

\bibitem[Rivera(2024)]{rivera2024situ} 
Rivera, YJ, Badman, ST, Stevens, ML, Verniero, JL, Stawarz, JE, Shi, C, Raines, JM, Paulson, KW, Owen, CJ, Niembro, T, et al. (2024). In situ observations of large-amplitude Alfv{\'e}n waves heating and accelerating the solar wind. *Science*, 385(6712), 962--966.

\bibitem[Zank(2022)]{zank2022turbulence} 
Zank, GP, Zhao, LL, Adhikari, L, Telloni, D, Kasper, JC, Stevens, M, Rahmati, A, Bale, SD. (2022). Turbulence in the sub-Alfv{\'e}nic solar wind. *The Astrophysical Journal Letters*, 926(2), L16.

\bibitem[Zank(2024)]{zank2024characterization} 
Zank, GP, Zhao, LL, Adhikari, L, Telloni, D, Baruwal, P, Baruwal, P, Zhu, X, Nakanotani, M, Pit{\v{n}}a, A, Kasper, JC, et al. (2024). Characterization of Turbulent Fluctuations in the Sub-Alfv{\'e}nic Solar Wind. *The Astrophysical Journal*, 966(1), 75.

\bibitem[Zank(2023)]{zank2023linear} 
Zank, GP, Zhao, LL, Adhikari, L, Nakanotani, M, Pit{\v{n}}a, A, Telloni, D, Che, H. (2023). Linear Mode Decomposition in Magnetohydrodynamics Revisited. *The Astrophysical Journal Supplement Series*, 268(1), 18.

\bibitem[Huang(2018)]{huang2018kinetic} 
Huang, H, Yu, Y, Dai, L, Wang, T. (2018). Kinetic Alfv{\'e}n waves excited in two-dimensional magnetic reconnection. *Journal of Geophysical Research: Space Physics*, 123(8), 6655--6669.

\bibitem[Hollweg(1978)]{hollweg1978alfven} 
Hollweg, JV. (1978). Alfv{\'e}n waves in the solar atmosphere. *Solar Physics*, 56, 305--333.

\bibitem[Hollweg(1981)]{hollweg1981alfven} 
Hollweg, JV. (1981). Alfv{\'e}n waves in the solar atmosphere: II: Open and closed magnetic flux tubes. *Solar Physics*, 70, 25--66.

\bibitem[Schwartz(1984)]{schwartz1984chromospheric} 
Schwartz, SJ, Cally, PS, Bel, N. (1984). Chromospheric and coronal Alfv{\'e}nic oscillations in non-vertical magnetic fields. *Solar Physics*, 92, 81--98.

\bibitem[Heyvaerts(1983)]{heyvaerts1983coronal} 
Heyvaerts, J, Priest, ER. (1983). Coronal heating by phase-mixed shear Alfven waves. *Astronomy and Astrophysics*, 117(2), 220--234.

\bibitem[Abdelatif(1987)]{abdelatif1987heating} 
Abdelatif, TE. (1987). Heating of coronal loops by phase-mixed shear Alfven waves. *Astrophysical Journal*, 322, 494--502.

\bibitem[Ionson(1978)]{ionson1978resonant} 
Ionson, JA. (1978). Resonant absorption of Alfv{\'e}nic surface waves and the heating of solar coronal loops. *Astrophysical Journal*, 226, 650--673.

\bibitem[Lee(1986)]{lee1986behavior} 
Lee, MA, Roberts, B. (1986). On the behavior of hydromagnetic surface waves. *Astrophysical Journal*, 301, 430--439.

\bibitem[Chin(1972)]{chin1972nonlinear} 
Chin, YC, Wentzel, DG. (1972). Nonlinear dissipation of Alfv{\'e}n waves. *Astrophysics and Space Science*, 16, 465--477.

\bibitem[Uchida(1974)]{uchida1974excess} 
Uchida, Y, Kaburaki, O. (1974). Excess heating of corona and chromosphere above magnetic regions by non-linear Alfv{\'e}n waves. *Solar Physics*, 35, 451--466.

\bibitem[Wentzel(1974)]{wentzel1974coronal} 
Wentzel, DG. (1974). Coronal heating by Alfv{\'e}n waves. *Solar Physics*, 39(1), 129--140.

\bibitem[Stefant(1970)]{stefant1970alfven} 
St{\'e}fant, RJ. (1970). Alfv{\'e}n wave damping from finite gyroradius coupling to the ion acoustic mode. *The Physics of Fluids*, 13(2), 440--450.

\bibitem[Hollweg(1971)]{hollweg1971density}
Hollweg, J. V. (1971). Density fluctuations driven by Alfvén waves. \textit{Journal of Geophysical Research}, 76(22), 5155--5161. Wiley Online Library.

\bibitem[Singh(2019)]{singh2019anisotropic}
Singh, H. D., \& Jatav, B. S. (2019). Anisotropic turbulence of kinetic Alfvén waves and heating in solar corona. \textit{Research in Astronomy and Astrophysics}, 19(12), 185. IOP Publishing.

\bibitem[Sahraoui(2009)]{sahraoui2009evidence}
Sahraoui, F., Goldstein, M. L., Robert, P., \& Khotyaintsev, Y. V. (2009). Evidence of a cascade and dissipation of solar-wind turbulence at the electron gyroscale. \textit{Physical review letters}, 102(23), 231102. APS.

\bibitem[Alexandrova(2012)]{alexandrova2012solar}
Alexandrova, O., Lacombe, C., Mangeney, A., Grappin, R., \& Maksimovic, M. (2012). Solar wind turbulent spectrum at plasma kinetic scales. \textit{The Astrophysical Journal}, 760(2), 121. IOP Publishing.

\bibitem[Alexandrova(2013)]{alexandrova2013solar}
Alexandrova, O., Chen, C. H. K., Sorriso-Valvo, L., Horbury, T. S., \& Bale, S. D. (2013). Solar wind turbulence and the role of ion instabilities. \textit{Space Science Reviews}, 178(2), 101--139. Springer.

\bibitem[Chen(2019)]{chen2019evidence}
Chen, C. H. K., Klein, K. G., \& Howes, G. G. (2019). Evidence for electron Landau damping in space plasma turbulence. \textit{Nature communications}, 10(1), 740. Nature Publishing Group UK London.

\bibitem[Parashar(2018)]{parashar2018dependence}
Parashar, T. N., Matthaeus, W. H., \& Shay, M. A. (2018). Dependence of kinetic plasma turbulence on plasma $\beta$. \textit{The Astrophysical Journal Letters}, 864(1), L21. IOP Publishing.

\bibitem[Hui(1992)]{hui1992electron}
Hui, C.-H., \& Seyler, C. E. (1992). Electron acceleration by Alfvén waves in the magnetosphere. \textit{Journal of Geophysical Research: Space Physics}, 97(A4), 3953--3963. Wiley Online Library.

\bibitem[Boehm(1990)]{boehm1990high}
Boehm, M. H., Carlson, C. W., McFadden, J. P., Clemmons, J. H., \& Mozer, F. S. (1990). High-resolution sounding rocket observations of large-amplitude Alfvén waves. \textit{Journal of Geophysical Research: Space Physics}, 95(A8), 12157--12171. Wiley Online Library.

\bibitem[Louarn(1994)]{louarn1994observation}
Louarn, P., Wahlund, J. E., Chust, T., De Feraudy, H., Roux, A., Holback, B., Dovner, P. O., Eriksson, A. I., \& Holmgren, G. (1994). Observation of kinetic Alfvén waves by the FREJA spacecraft. \textit{Geophysical Research Letters}, 21(17), 1847--1850. Wiley Online Library.

\bibitem[Boehm(1995)]{boehm1995freja}
Boehm, M. H., Clemmons, J., \& Paschmann, G. (1995). Freja observations of a ten-meter boundary within monoenergetic auroral electron precipitation. \textit{Geophysical research letters}, 22(1), 69--72. Wiley Online Library.

\bibitem[Chaston(2014)]{chaston2014heating}
Chaston, C. C., Bonnell, J. W., \& Salem, C. (2014). Heating of the plasma sheet by broadband electromagnetic waves. \textit{Geophysical Research Letters}, 41(23), 8185--8192. Wiley Online Library.

\bibitem[Kerr(2016)]{kerr2016simulations}
Kerr, G. S., Fletcher, L., Russell, A. J. B., \& Allred, J. C. (2016). Simulations of the Mg II k and Ca II 8542 lines from an Alfvén Wave-heated Flare Chromosphere. \textit{The Astrophysical Journal}, 827(2), 101. IOP Publishing.

\bibitem[Reep(2016)]{reep2016alfvenic}
Reep, J. W., \& Russell, A. J. B. (2016). Alfvénic wave heating of the upper chromosphere in flares. \textit{The Astrophysical Journal Letters}, 818(1), L20. IOP Publishing.

\bibitem[Reep(2018)]{reep2018hydrodynamic}
Reep, J. W., Russell, A. J. B., Tarr, L. A., \& Leake, J. E. (2018). A hydrodynamic model of Alfvénic wave heating in a coronal loop and its chromospheric footpoints. \textit{The Astrophysical Journal}, 853(2), 101. IOP Publishing.

\bibitem[Artemyev(2015)]{artemyev2015acceleration}
Artemyev, A. V., Liu, J., Angelopoulos, V., \& Runov, A. (2015). Acceleration of ions by electric field pulses in the inner magnetosphere. \textit{Journal of Geophysical Research: Space Physics}, 120(6), 4628--4640. Wiley Online Library.

\bibitem[Damiano(2015)]{damiano2015ion}
Damiano, P. A., Johnson, J. R., \& Chaston, C. C. (2015). Ion temperature effects on magnetotail Alfvén wave propagation and electron energization. \textit{Journal of Geophysical Research: Space Physics}, 120(7), 5623--5632. Wiley Online Library.

\bibitem[Damiano(2016)]{damiano2016ion}
Damiano, P. A., Johnson, J. R., \& Chaston, C. C. (2016). Ion gyroradius effects on particle trapping in kinetic Alfvén waves along auroral field lines. \textit{Journal of Geophysical Research: Space Physics}, 121(11), 10--831. Wiley Online Library.

\bibitem[Parashar(2015)]{parashar2015transition}
Parashar, T. N., Matthaeus, W. H., Shay, M. A., \& Wan, M. (2015). Transition from kinetic to MHD behavior in a collisionless plasma. \textit{The Astrophysical Journal}, 811(2), 112. IOP Publishing.

\bibitem[Ayaz(2024b)]{ayaz2024alfven}
Ayaz, S., Zank, G. P., Khan, I. A., et al. (2024). Alfvén waves in the solar corona: resonance velocity, damping length, and charged particles acceleration by kinetic Alfvén waves. \textit{Scientific Reports}, 14, 27275. https://doi.org/10.1038/s41598-024-77975-6.

\end{thebibliography}
%

\begin{appendix}\label{Appendix: A}
\section{Background mathematical formalism}
The distribution function $f_{0s}(v)$ given in Eq. (\ref{eq:2}) is used to find out the permittivity tensor components ($\epsilon_{xx}$ and $\epsilon_{zz}$) in 
\begin{equation}
    \mathrm{D}=(\epsilon_{xx}-n_{\parallel}^2)\epsilon_{zz}-\epsilon_{xx}n_{\perp}^2
    \label{akk},
\end{equation}
which are related to the distribution function through
\begin{equation}
    \epsilon_{xx}=1+\sum_{s}\frac{\omega_{ps}^2}{\omega}\int d^3 v\sum_{n=-\infty}^{\infty}\frac{n^2}{\zeta_{s}^2}\frac{v_{\perp}J_{n}^2(\zeta_s)}{(\omega-k_{\parallel}v_{\parallel}-n\Omega_{s})}\frac{\partial f_{0s}(v)}{\partial v_{\perp}} 
    \label{eq:3},
\end{equation}
and \begin{equation}
    \epsilon_{zz}=1+\sum_{s}\frac{\omega_{ps}^2}{\omega}\int d^3 v\sum_{n=-\infty}^{\infty}\frac{J_{n}^2(\zeta_s)}{(\omega-k_{\parallel}v_{\parallel}-n\Omega_{s})}\frac{v_{\parallel}^2}{v_{\perp}}\frac{\partial f_{0s}(v)}{\partial v_{\perp}}
    \label{eq:4}.
\end{equation}
Here, the plasma and gyro frequencies are $\omega_{ps}=\sqrt{4\pi n_{0}e^2/m_{s}}$ and $\Omega_{s}=q_{s}\mathrm{B}_{0}/m_{s}c$, respectively. $J_{n}(\zeta_s)$ is the Bessel function with an argument $\zeta_s=k_{\perp}v_{\perp}/\Omega_{s}$. In Eqs. (\ref{eq:3}) and (\ref{eq:4}), using $f_{0s}(v)$ and solving the parallel and perpendicular integrals with the assumption that the ion gyroradius is smaller than the perpendicular wavelength (i.e., $k_{\perp}\rho_{i}\ll 1$), to get \citep{Khan2019a, khan2020solar}
\begin{equation}
    \epsilon_{xx}=\frac{c^2}{\mathrm{v_{A}}^2}\Bigl(1-\frac{3}{4}\frac{2\kappa}{2\kappa-3}k_{\perp}^2\rho_{i}^2\Bigl) 
    \label{eq:5}
\end{equation}
and 
\begin{equation}
    \epsilon_{zz}=\frac{2\omega_{pe}^2}{k_{\parallel}^2 v_{Te}^2} \frac{2\kappa-1}{2\kappa}+2i\frac{\omega_{pe}^2 \omega_{r}}{k_{\parallel}^3 v_{Te}^3}\frac{1}{\kappa^{3/2}}\frac{\Gamma(\kappa+1)}{\Gamma(\kappa-1/2)}\sqrt{\pi}  \label{eq:6},
\end{equation} 
where $\mathrm{v_{A}}=\frac{\mathrm{B}_{0}}{\sqrt{4\pi n_{0}m_{i}}}$ is Alfvénic speed and $\rho_{i}^2(=\mathrm{v_{Ti}^2}/2\Omega_{i}^2$) is the ion gyroradius.

Use Eqs. (\ref{eq:5}) and (\ref{eq:6}) in Eq. (\ref{akk}) and take $\mathrm{D}=0$ to obtain the real and imaginary frequencies:
\begin{equation}
\omega_{r}=k_{\parallel} \mathrm{v_{A}}\Biggr[\Bigr(1+\frac{3}{4}\frac{2\kappa}{2\kappa-3}k_{\perp}^2 \rho_{i}^2\Bigr)+\frac{2\kappa}{2\kappa-1}\mathrm{\frac{T_e}{T_i}}k_{\perp}^2 \rho_{i}^2\Biggr]^{1/2}
\label{eq:7},
\end{equation}
and 
\begin{equation}
   \omega_i=-k_{\parallel}\mathrm{v_{A}}\gamma 
   \label{eq:8},
\end{equation}
with \begin{equation}
\begin{split}
     \gamma=\frac{\mathrm{v_{A}}}{\mathrm{v_{T e}}}\Biggr(1+\frac{3}{4}\frac{2\kappa}{2\kappa-1}\mathrm{\frac{T_e}{T_i}}k_{\perp}^2 \rho_{i}^2\Biggr)^{-1}\\
     \times \Biggr[\mathrm{\frac{T_e}{T_i}}k_{\perp}^2 \rho_{i}^2 \frac{2\kappa^2}{(2\kappa-1)^2}\frac{\Gamma(\kappa+1)}{\Gamma(\kappa-1/2)}\sqrt{\pi}\Biggr]\\
     \times \Biggl\{1+\sqrt{\frac{m_i}{m_e}}\Bigr(\mathrm{\frac{T_e}{T_i}}\Bigr)^{3/2}\Biggr(1+\frac{1}{\kappa}\frac{\omega_{r}^2}{k_{\parallel}^2 \mathrm{v_{Ti}^2}}\Biggr)^{-\kappa-1}\Biggl\}
    \label{eq:9}.
    \end{split}
\end{equation}
The above expressions, particularly Eq. (\ref{eq:9}), illustrate the role of Landau damping. This equation, however, doesn't explicitly illustrate how EM energy is converted into heat as the waves propagate through space. To understand this transformation, we must explore the relationship between the Poynting flux vector and the heating rate. The following sections explore this crucial element, providing a comprehensive analysis.

\subsection{Derivation of the perturbed EM fields}
Starting from the simple Faraday's law
\begin{equation*}
\nabla \times \mathbf{E}=-\frac{\partial \mathbf{B}}{\partial t}
\end{equation*}
translates into
\begin{equation}
\mathbf{k}\times \mathbf{E}=\omega \mathbf{B}\label
{eq:10},
\end{equation} 
where we replaced $\nabla $ by $i\mathbf{k}$ and $\frac{\partial}{\partial t}$ by $-i\omega $. As we assumed that $\textbf{k}$ and $\textbf{E}$ lie in the x-z plane and the $\textbf{B}$ of the wave is along the y axis, Eq. (\ref{eq:10}) can also be written as
\begin{equation}
    \mathrm{\frac{E_x}{B_y}}=\frac{\omega}{k_{\parallel}-k_{\perp}\mathrm{E_{z}/E_{x}}} \label{eq:11}.
\end{equation}
In Eq. (\ref{eq:11}), the ratio $\mathrm{E_{z}/E_{x}}$ in the denominator can be obtained by solving the first row of Eq. (\ref{eq:1}). Thus,
\begin{equation}
 \mathrm{\frac{E_z}{E_x}}=\frac{c^2 k_{\parallel}^2 -\omega^2 \epsilon_{xx}}{c^2 k_{\parallel}k_{\perp}}    
 \label{eq:12}.
\end{equation}
Using Eq. (\ref{eq:12}) and Eqs. (\ref{eq:5} - \ref{eq:9}) with the assumption on $\omega=\omega_{r}+i\omega_i$, we get
\begin{equation}
    \mathrm{\frac{E_x}{B_y}}=\frac{c^2 k_{\parallel}}{\epsilon_{xx}}\Bigr(\frac{\omega_r}{\omega_r^{2}+\omega_i^2}-i\frac{\omega_i}{\omega_r^{2}+\omega_i^2}\Bigr)
    \label{eq:13}.
\end{equation}
In the assumption that $\omega_r\gg \omega_i$, the real and imaginary parts in Eq. (\ref{eq:13}) yields
\begin{equation}
    \text{Re} \Bigr(\mathrm{\frac{E_x}{v_A B_y}}\Bigr)=\frac{\Bigr(1+\frac{3}{4}\frac{2\kappa}{2\kappa-3}k_{\perp}^2 \rho_{i}^2\Bigr)}{\Bigr[\Bigr(1+\frac{3}{4}\frac{2\kappa}{2\kappa-3}k_{\perp}^2 \rho_{i}^2\Bigr)+\frac{2\kappa}{2\kappa-1}\mathrm{\frac{T_e}{T_i}}k_{\perp}^2 \rho_{i}^2\Bigr]^{1/2}}
    \label{eq:14}
\end{equation}
and 
\begin{equation}
    \text{Im} \Bigr(\mathrm{\frac{E_x}{v_A B_y}}\Bigr)=\frac{\mathrm{v_{A}}}{\mathrm{v_{Te}}}C_k\times\Biggl\{1+\sqrt{\frac{m_i}{m_e}}\Bigr(\mathrm{\frac{T_e}{T_i}}\Bigr)^{3/2}\Biggr(1+\frac{1}{\kappa}\frac{\omega_{r}^2}{k_{\parallel}^2 \mathrm{v_{Ti}^2}}\Biggr)^{-\kappa-1}\Biggl\}\sqrt{\pi}
    \label{eq:15}.
\end{equation}
Here, we define $C_k=\Biggr[\frac{(1+\frac{3}{4}\frac{2\kappa}{2\kappa-3}k_{\perp}^2 \rho_{i}^2)\Bigr(\mathrm{\frac{T_e}{T_i}}k_{\perp}^2 \rho_{i}^2 \frac{2\kappa^2}{(2\kappa-1)^2}\frac{\Gamma(\kappa+1)}{\Gamma(\kappa-1/2)}\Bigr)}{(1+\frac{3}{4}\frac{2\kappa}{2\kappa-1}\mathrm{\frac{T_e}{T_i}}k_{\perp}^2 \rho_{i}^2)\Bigr(1+\frac{3}{4}\frac{2\kappa}{2\kappa-3}k_{\perp}^2 \rho_{i}^2 +\frac{2\kappa}{2\kappa-1}\mathrm{\frac{T_e}{T_i}}k_{\perp}^2 \rho_{i}^2\Bigr)}\Biggr]$.

Eq. (\ref{eq:15}) has often been overlooked due to its minimal contributions in certain regions of space, such as the auroral region \citep{Lysak2003,lysak2023kinetic} and magnetopause \citep{khan2019distinct}. However, in our study of KAWs in the solar corona, even these small contributions are crucial. Therefore, we emphasize the importance of not disregarding the minor components of the imaginary perturbed EM field ratio in our analysis.

Using the same algebra as previously, Eq. (\ref{eq:12}) gives
\begin{equation}
    \mathrm{\frac{E_z}{E_x}}=\frac{c^2 k_{\parallel}^2-\Bigr(\omega_r^2+2i\omega_r \omega_i\Bigr)\frac{c^2}{\mathrm{v_{A}}^2}\Bigr(1-\frac{3}{4}\frac{2\kappa}{2\kappa-3}k_{\perp}^2 \rho_i^2\Bigr)}{c^2 k_{\parallel}k_{\perp}}
    \label{eq:16}.
\end{equation}
The real and imaginary parts in Eq. (\ref{eq:16}) are
\begin{equation}
\begin{split}
    \frac{k_{\perp}}{k_{\parallel}} \text{Re}\Bigr(\mathrm{\frac{E_z}{E_x}}\Bigr)= \\
    \Biggr[1-\Bigr(1+\frac{3}{4}\frac{2\kappa}{2\kappa-3}k_{\perp}^2 \rho_i^2+\frac{2\kappa}{2\kappa-1}\mathrm{\frac{T_e}{T_i}}k_{\perp}^2 \rho_i^2\Bigr)\times \Bigr(1-\frac{3}{4}\frac{2\kappa}{2\kappa-3}k_{\perp}^2 \rho_i^2\Bigr)\Biggr]
    \label{eq:17}
    \end{split}
\end{equation}
and 
\begin{equation}
\begin{split}
    \frac{k_{\perp}}{k_{\parallel}}\text{Im}\Bigr(\mathrm{\frac{E_z}{E_x}}\Bigr)=2\gamma \Bigr(1-\frac{3}{4}\frac{2\kappa}{2\kappa-3}k_{\perp}^2 \rho_i^2\Bigr)\\
    \times \Biggr[\Bigr(1+\frac{3}{4}\frac{2\kappa}{2\kappa-3}k_{\perp}^2 \rho_i^2\Bigr)+\frac{2\kappa}{2\kappa-1}\mathrm{\frac{T_e}{T_i}}k_{\perp}^2 \rho_i^2\Biggr]^{1/2}
    \label{eq:18},
    \end{split}
\end{equation}
respectively.

\section{Derivations of the Poynting flux vectors, power rate, resonant velocity, group speed, and the damping length expressions}\label{Appendix: B}

\subsection{The Poynting flux and heating rate of KAWs}

We invoke the simplified EM field ratios (i.e., Eqs. \ref{eq:11} - \ref{eq:13}) to find the Poynting flux vector of KAWs. On averaging over a complete cycle, the differential term in the Poynting theorem 
\begin{equation}
    \frac{\partial u}{\partial t}=-\textbf{J}\cdot \textbf{E}-\nabla\cdot \mathbf{S}
    \label{eq:19}
\end{equation}
\citep{Lysak2003} is zero for sinusoidal EM perturbations. Simply put, we can write
\begin{equation}
    \nabla \cdot \mathbf{S}=-\mathrm{P}
    \label{eq:20},
\end{equation}
with $\mathbf{S}$ being the Poynting flux vector given by
\begin{equation*}
    \mathbf{S}=\text{Re} \Bigr(\mathrm{E}^{\ast}\times \textbf{B}\Bigr)/2\mu_0,
\end{equation*}
and $\mathrm{P}$ the power dissipation, given as
\begin{equation*}
   \mathrm{P}=\text{Re} \Bigr(\mathrm{J}^{\ast}\cdot \textbf{E}\Bigr)/2.
\end{equation*}
In the above expressions, $\mu_0$ represents the permeability of free space, and $\mathbf{J}$ is the current density, which can be determined using Ampere's law: $\mathbf{J} = \left(\nabla \times \textbf{B}\right)/\mu_0$. The factor of 1/2 arises from averaging over a complete cycle.

As is shown in Fig. (\ref{A}), the wave is propagating in the x-z plane, therefore, the y-component of the Poynting flux vector ($\mathrm{S_y}$) is naturally zero. Following \cite{Ayaz2024}, we can write the Poynting flux vector of KAWs as 
\begin{equation}
\nabla \cdot \mathbf{S}=\frac{\partial}{\partial x}\mathrm{S_{x} (x,z)}+\frac{\partial}{\partial z}\mathrm{S_{z} (x,z)}  
\label{eq:21}.
\end{equation}
In Eq. (\ref{eq:21}), the first term represents the Poynting flux in the perpendicular direction and the second term gives the Poynting flux in the parallel direction. The averaged x and z components of the Poynting vectors alone in the above expression are
\begin{equation}
    \mathrm{S_x}=-\Bigl(\mathrm{\frac{E_z}{E_x}\Bigl)S_z}
    \label{eq:22}
\end{equation}
\citep{Ayaz2024} and 
\begin{equation}
    \mathrm{S_z= \text{Re} (E^*_x B_y)} 
    \label{eq:23}
\end{equation}
\citep{Lysak2003}, respectively.

In general, KAWs, characterized by $k_{\perp}\gg k_{\parallel}$ \citep{Narita2020}, can carry some energy in the perpendicular direction (i.e., across the field lines), but that energy is typically small \citep{Lysak2003, khan2020solar}. However, the perpendicular energy transport of KAWs, defined by the $\mathrm{S_x}$ expression, is often overlooked in certain environments, such as plasma sheaths, magnetopause, and auroral zones. Here, we focus on the solar coronal region and recognize the significance of even this small $\mathrm{S_x}$ contribution. This is crucial for fully understanding the energy transport mechanisms of KAWs in solar flux loop tubes, as is shown in Fig. \ref{A} (right panel). 

The parallel Poynting flux vector $\mathrm{S_z}$ can be simplified using the power dissipation expression
\begin{equation}
   \mathrm{P}=\text{Re} \Bigr(\mathrm{J}^{\ast}\cdot \textbf{E}\Bigr)/2=k_{\perp} \text{Re} \Bigr(\mathrm{\frac{E_z}{E_x}}\Bigr)\mathrm{S_z}=2\gamma k_{\parallel}\mathrm{S_z} R
   \label{eq:24},
\end{equation}
with $R=\Bigr(1+\frac{3}{4}\frac{2\kappa}{2\kappa-3}k_{\perp}^2 \rho_i^{2}\Bigr)\Bigr[\Bigr(1+\frac{3}{4}\frac{2\kappa}{2\kappa-3}k_{\perp}^2 \rho_i^{2}+\frac{2\kappa}{2\kappa-1}\mathrm{\frac{T_e}{T_i}}k_{\perp}^2 \rho_i^{2}\Bigr)\Bigr]$.

For the z-component, the steady-state form of the Poynting flux theorem requires:
\begin{equation}
    \frac{\partial \mathrm{S_z} (z)}{\partial z}=-\mathrm{P}=2\gamma k_{\parallel}\mathrm{S_z} R
    \label{eq:25}.
\end{equation}
The solution to the above equation is
\begin{equation}
    \mathrm{S_z (z)=S_z (0)}\times e^{-2\gamma k_{\parallel}\mathrm{z} R}
    \label{eq:26},
\end{equation}
which describes the Poynting flux of KAWs in the parallel direction, highlighting how the waves transport energy as they propagate in space.

Similarly, substituting Eq. (\ref{eq:26}) in Eq. (\ref{eq:22}), we obtain the solution for $\mathrm{S_x}$ as
\begin{equation}
  \mathrm{S_z (x)}=-\Bigr(\mathrm{\frac{E_z}{E_x}}\Bigr)\mathrm{S_z (0)}\times e^{-2\gamma k_{\parallel}\mathrm{z} R}
    \label{eq:27}.  
\end{equation}
Eqs. (\ref{eq:26}) and (\ref{eq:27}) provide a framework to quantify the conversion of EM energy into thermal energy within solar flux tube loops as the waves propagate from the initial reference point z = 0. At z = 0, where the waves are first excited, the Poynting flux has an initial magnitude of $\mathrm{S (0)}$. As these waves travel through the plasma, the Poynting flux decreases due to the negative imaginary part of the wave frequency, $\omega_i$; that is, energy transfer from the waves to the plasma particles. Physically, this phenomenon results from wave-particle interactions, where plasma particles, including ions and electrons, absorb energy from the waves. This energy transfer causes the particles to gain kinetic energy, leading to plasma heating. The process is particularly significant in the solar corona, where such interactions can contribute to the high temperatures observed.

Solar flux tube loops, which are magnetic structures extending from the solar surface into the corona, play a critical role in energy transport and heating within the solar atmosphere. The plasma within a coronal loop is typically higher than the plasma outside. The complex processes occurring in these regions, particularly the mechanisms related to coronal heating including the turbulent cascade to very small dissipation length scales or in general kinetic scales \citep{marsch2006kinetic,browning1991mechanisms}. Coronal heating remains an area of active research and debate.

Several seminal works, including those by \cite{matthaeus1999coronal}, and \cite{cranmer2010can}, have laid the foundation for understanding the turbulent cascade in the solar corona. These studies describe how energy is transferred to progressively smaller perpendicular scales until it dissipates, typically at ion inertial or gyrofrequency scales. More recent investigations by \cite{adhikari2024turbulence}, \cite{zank2018theory, zank2021turbulence}, and \cite{yalim2024mixing} continue to explore the intricacies of solar coronal heating, shedding light on the complex interplay of turbulence and wave dissipation.

In particular, within coronal loops, Alfv\'{e}n waves and quasi 2D fluctuations are generated by dynamic twisting and braiding motions at the loop's footpoints on the photosphere \citep{zank2018pickup}. These motions, caused by convective flows in intergranular lanes, distort magnetic flux tubes and create Alfv\'{e}n waves that propagate outward along the magnetic field lines advected in 2D turbulence. As Alfv\'{e}n waves travel through the chromosphere and corona, they dissipate their energy, contributing to the heating of the solar atmosphere. This process, while better understood today, continues to be a focal point of research, especially concerning the behavior of Alfv\'{e}n waves (i.e., KAWs) at small kinetic scales and their role in turbulent energy cascade and dissipation.

\subsection{The net power deposition by KAWs}

The power transfer rate of KAWs in solar flux tube loops is a crucial indicator of energy distribution and absorption in the solar corona. As the waves propagate, their Poynting flux dissipates due to wave-particle interactions, where plasma particles absorb energy from the waves, leading to heating. This energy transfer is essential for explaining the high temperatures observed in the solar corona. It plays a significant role in solar wind acceleration, magnetic reconnection, and solar flares. Here, we quantify the power transfer rate of KAWs.

To derive the power transfer rate of KAWs, we substitute $\mathrm{z=h\theta}$ in Eqs. (\ref{eq:26}) and (\ref{eq:27}) and integrate from $0$ to $\pi$ and $\pi/2$, to obtain
\begin{equation}
    \mathrm{I_x}=a\Bigr(\mathrm{\frac{E_z}{E_x}}\Bigr)\Biggr(\frac{-1+e^{-\gamma k_{\parallel}\mathrm{h}\pi R}}{\gamma k_{\parallel}R}\Biggr)\pi
    \label{eq:28},
\end{equation}
and 
\begin{equation}
    \mathrm{I_z}=a^2\Bigr(e^{-\gamma k_{\parallel}\mathrm{h}\pi R}\Bigr)\pi
    \label{eq:29}.
\end{equation}
In Eqs. (\ref{eq:28}) and (\ref{eq:29}), $a$ represents the crosssection of the semicircle, which ranges from $700 - 700$0 km \citep{li2023modeling}, and h is the height of the flux tube loop. \cite{Effenberger2017} examined the height of $61$ occulted flare loops, and we follow that study by assuming an h range of $0.05 - 0.1$ R\textsubscript{Sun}. The same range was considered in \cite{li2023modeling} to examine solar energetic neutral particles. By taking the ratio of Eqs. (\ref{eq:28}) and (\ref{eq:29}), the total power transfer rate emerging from the tube loop is
\begin{equation}
    \frac{\mathrm{I_x}}{\mathrm{I_z}}=-\Bigr(\mathrm{\frac{E_z}{E_x}}\Bigr)\Biggr(\frac{1-e^{-\gamma k_{\parallel}\mathbf{h}\pi R}}{a \gamma k_{\parallel}R e^{-\gamma k_{\parallel}\mathrm{h}\pi R}}\Biggr)
    \label{eq:30}.
\end{equation}

\subsection{The resultant resonance velocity of the particles}
Following \cite{paraschiv2015physical}, the kinetic energy flux passing is expressed as
\begin{equation}
    \mathrm{K} \cdot \mathrm{E}=\frac{1}{2}\rho \mathrm{v^3}
    \label{eq:31},
\end{equation}
with $\rho$ being the mass density and $\mathrm{v}$ be the speed of the particles, which can be written in terms of the Poynting flux vector $\mathrm{(S_{x,z})}$ as
\begin{equation}
    \mathrm{S_{x,z}}=\frac{1}{2}\rho v^3
    \label{eq:32},
\end{equation}
which gives
\begin{equation}
    \mathrm{v}=\Bigr(\frac{2\mathrm{S_{x,z}}}{\rho}\Bigr)^{1/3}
    \label{eq:33}.
\end{equation}
Eq. (\ref{eq:33}) represents the particle's speed -- the particles gained from the wave.

In general, when the particles interact with the wave, they receive energy from the wave to fulfill the resonance condition. In this situation, the wave delivers its energy to particles and is damped. The particles now have a net velocity that must be equal to the particle's initial velocity and the energy that the particle gains from the wave giving
\begin{equation}
    \mathrm{v_{net}}=\frac{\omega_r}{k_{\parallel}}+\Bigr(\frac{2\mathrm{S_{x,z}}}{\rho}\Bigr)^{1/3}
    \label{eq:34}.
\end{equation}
The first term in Eq. (\ref{eq:34}) represents the wave phase velocity at which particles resonate with the wave, while the second term quantifies the energy gained by the particles from the wave. Physically, this can be visualized as the wave giving a push to the particles under resonance conditions, resulting in an overall higher velocity for the particles after Landau resonance occurs. Consequently, the particle has a high speed and hence increased energy.

On substituting Eqs. (\ref{eq:7}) and (\ref{eq:26}) in Eq. (\ref{eq:34}), we obtain the expression for the normalized parallel speed of the particles as
\begin{equation}
\begin{split}
    \mathrm{\frac{v_{net}}{v_A}}=\Biggr[\Bigr(1+\frac{3}{4}\frac{2\kappa}{2\kappa-3}k_{\perp}^2\rho_{i}^2\Bigr)+\frac{2\kappa}{2\kappa-1}\mathrm{\frac{T_e}{T_i}} k_{\perp}^2\rho_{i}^2\Biggr]^{1/2}\\
    +\frac{1}{\mathrm{v_A}}\Biggr(\frac{2\mathrm{S_{z}(0)}\times e^{-2\gamma k_{\parallel}\mathrm{z} R}}{\rho}\Biggr)^{1/3}
    \label{eq:35}.
    \end{split}
\end{equation}
Similarly, employing Eqs. (\ref{eq:7}) and (\ref{eq:27}) in Eq. (\ref{eq:34}) to obtain the normalized perpendicular velocity expression as
\begin{equation}
\begin{split}
    \mathrm{\frac{v_{net\perp}}{v_A}}=\Biggr[\Bigr(1+\frac{3}{4}\frac{2\kappa}{2\kappa-3}k_{\perp}^2\rho_{i}^2\Bigr)+\frac{2\kappa}{2\kappa-1}\mathrm{\frac{T_e}{T_i}}k_{\perp}^2\rho_{i}^2\Biggr]^{1/2}\\
    +\frac{1}{\mathrm{v_A}}\Biggr[-\Bigr(\mathrm{\frac{E_z}{E_x}}\Bigr)\frac{2\mathrm{S_{z}(0)}\times e^{-2\gamma k_{\parallel}\mathrm{z}R}}{\rho}\Biggr]^{1/3}
    \label{eq:36}.
    \end{split}
\end{equation}
The above expressions (\ref{eq:35}) and (\ref{eq:36}) yield the net parallel and perpendicular velocity for the particles experiencing Landau damping in the solar corona. During this interaction, particles gain energy from the waves while the wave itself experiences damping.

In the parallel direction (i.e., Eq. (\ref{eq:35})), the net velocity is boosted due to the efficient transfer of wave energy along the magnetic field lines, which is also shown in the numerical analysis in the later section. This interaction primarily heats and accelerates the particles parallel to the ambient magnetic field. Similarly, although the contribution is generally smaller in the perpendicular direction (i.e., Eq. (\ref{eq:36})), the particles still gain velocity due to wave-particle interactions, leading to perpendicular heating and acceleration. These dual mechanisms of energy transfer are crucial for understanding the dynamics within the solar corona.

This scenario highlights the significant role of KAWs in plasma heating and particle acceleration in the solar corona. The derived expressions for net velocity in both directions underscore how KAWs can contribute to the high temperatures observed in the corona and the efficient energy distribution throughout the solar atmosphere. This process not only elucidates the behavior of particles in the solar corona but also provides insights into broader astrophysical phenomena where similar wave-particle interactions are at play.

\subsection{The group velocity and the characteristic damping length of KAWs}

Following \cite{tiwari2008shear}, the damping length expression is
\begin{equation}
    \mathrm{L_G}=\frac{\mathrm{v_G}}{\omega_i}
    \label{eq:37},
    \end{equation}
where $\mathrm{v_G}$ is the group velocity defined as:
\begin{equation}
    \mathrm{v_G}=\sqrt{\Bigr(\frac{\partial \omega_r}{\partial k_{\perp}}\Bigr)^2+\Bigr(\frac{\partial \omega_r}{\partial k_{\parallel}}\Bigr)^2} 
    \label{eq:38}.
\end{equation}
On evaluating Eq. (\ref{eq:7}) and substituting those values in Eq. (\ref{eq:38}), we obtain the normalized group velocity as
\begin{equation}
    \frac{\mathrm{v_G}}{\mathrm{v_{A}}}=\Biggr[Q+\Bigr(1+\frac{3}{4}\frac{2\kappa}{2\kappa-3}k_{\perp}^2 \rho_{i}^2\Bigr)+\frac{2\kappa}{2\kappa-1}\mathrm{\frac{T_e}{T_i}}k_{\perp}^2 \rho_{i}^2\Biggr]^{1/2}
    \label{eq:39}.
\end{equation}
Here we define, $Q=\frac{k_{\parallel}^2 \Bigr[(\frac{3}{4}\frac{2\kappa}{2\kappa-3}k_{\perp}\rho_{i}^2)+\frac{2\kappa}{2\kappa-1}\mathrm{\frac{T_e}{T_i}}k_{\perp}\rho_{i}^2\Bigr]^2}{(1+\frac{3}{4}\frac{2\kappa}{2\kappa-3}k_{\perp}^2\rho_{i}^2)+\frac{2\kappa}{2\kappa-1}\mathrm{\frac{T_e}{T_i}}k_{\perp}^2\rho_{i}^2}$.

Use of Eqs. (\ref{eq:8}) and (\ref{eq:39}) in (\ref{eq:37}) yields
\begin{equation}
    \mathrm{L_G}=\frac{\mathrm{v_{Te}}\Bigr(1+\frac{3}{4}\frac{2\kappa}{2\kappa-1}\mathrm{\frac{T_e}{T_i}}k_{\perp}^2\rho_{i}^2\Bigr)\Biggr[Q+\Bigr(1+\frac{3}{4}\frac{2\kappa}{2\kappa-3}k_{\perp}^2\rho_{i}^2\Bigr)+\frac{2\kappa}{2\kappa-1}\mathrm{\frac{T_e}{T_i}}k_{\perp}^2\rho_{i}^2\Biggr]^{1/2}}{\Bigr[\mathrm{\frac{T_e}{T_i}}k_{\perp}^2\rho_{i}^2 \sqrt{\pi} \frac{2\kappa^2}{(2\kappa-1)^2}\frac{\Gamma(\kappa+1)}{\Gamma(\kappa-1/2)}\Bigr]\times\Biggr\{1+\sqrt{\frac{m_e}{m_i}}\Bigr(\mathrm{\frac{T_e}{T_i}}\Bigr)^{3/2}\Biggr(1+\frac{1}{\kappa}\frac{\omega_{r}^2}{k_{\parallel}^2 \mathrm{v_{Ti}^2}}\Biggr)^{-\kappa-1}\Biggr\}}
  \label{eq:40}.
\end{equation}
\end{appendix}
\end{document}